\documentclass[prc,nofootinbib,showpacs]{revtex4}

\usepackage{graphicx}
\usepackage[usenames,dvipsnames]{color}
\usepackage{amsmath,amssymb,bbold}
\usepackage{hyperref}
\usepackage{slashed}
\usepackage{comment}
\begin{document}

\title{Spin Polarization and Chiral Condensation in 2+1 flavor Nambu-Jona-Lasinio model
at finite temperature and baryon chemical potential}

\author{Aman Abhishek}
\email{aman@prl.res.in}
\affiliation{Theory Division, Physical Research Laboratory,
	Navrangpura, Ahmedabad 380 009, India}
\affiliation{Indian Institute of Technology Gandhinagar, Gandhinagar 382355, Gujarat, India}

\author{Arpan Das}
\email{arpan@prl.res.in}
\affiliation{Theory Division, Physical Research Laboratory,
	Navrangpura, Ahmedabad 380 009, India}
	
\author{Hiranmaya Mishra}
\email{hm@prl.res.in}
\affiliation{Theory Division, Physical Research Laboratory,
	Navrangpura, Ahmedabad 380 009, India}	
	
	\author{Ranjita K. Mohapatra}
\email{ranjita@iitb.ac.in}
\affiliation{Department of Physics, Indian Institute of Technology Bombay, Powai, Mumbai 400076, India}

\begin{abstract}
We investigate the ferromagnetic (spin polarization) condensation in (2+1) flavor Nambu Jona-Lasinio(NJL) model with non-zero current quark masses at finite temperature and 
density which may be relevant in the context of neutron stars. The spin polarization condensation arises due to a tensor type interaction which may be generated due to non-
perturbative effects in Quantum Chromodynamics(QCD). 
In this investigation we have shown the interplay between chiral condensate and spin polarization condensation for different values of 
tensor coupling. Spin polarization in the case of 2+1 flavor is different from two flavor case because of additional 
F$_8$ condensate associated with $\lambda_8^f$ flavor generator. We find a non-zero value of the two spin condensates
in the chirally restored phase. Beyond a certain temperature 
the spin polarization condensates vanish for rather large quark chemical potentials. The spin condensates affect the chiral 
phase transition, quark masses and the quark dispersion relation. The spin polarization condensate appear only in the chiral restored
phase for light quarks. For large enough tensor couplings, it is observed that the spin polarization condensate acts as a catalyst for chiral
symmetry restoration. Thermodynamic behavior of $F_3$ and $F_8$ are found to be different and they affect the quark masses differently. 
\end{abstract}

\maketitle

\section{INTRODUCTION}
\label{intro}

One of the recent interests in high energy physics is to study the phase diagram of strongly interacting matter.
QCD phase diagram has been studied extensively in the temperature $(T)$ - baryon chemical potential
$(\mu_B)$ plane \cite{rischke2004,rajawilczek}. The first principle lattice QCD (LQCD) simulations give a
reliable prediction about the nature of QCD phases and phase transitions at zero baryon chemical
potential and finite temperature \cite{karsch2002,laerman2003,cheng2008}. Although LQCD calculations can be
trusted undoubtedly at small baryon chemical potential $\mu_B\simeq0$, at relatively large baryon chemical
potential lattice calculations suffer from the ``fermion sign problem''\cite{foder2002}. LQCD calculations 
predict that at $\mu_B=0$, the nature of the transition from confined hadronic phase to deconfined quark gluon plasma (QGP) 
phase is not a thermodynamic phase transition rather it is a smooth crossover with a transition temperature
$T_c\in [149-163]$ MeV \cite{Ding2014}. On the other hand, QCD inspired effective field theory models e.g. Nambu-Jona-Lasinio model (NJL) etc.,
indicate that the phase transition from the hadronic phase to QGP phase at large baryon chemical potential
and small temperature is first order in nature with physical quark masses. This indicates the presence of a critical endpoint at the end of the first
order chiral phase transition line in the QCD phase diagram. Apart from the confined hadronic phase and deconfined QGP phases, 
QCD phase diagram has a very rich structure at a low temperature
and high baryon chemical potential. In this region of  the phase diagram, possibility of various  exotic
phases has been investigated such as the color superconducting phase\cite{Rajagopal2008,Rajagopal1999,byam2001},
quarkyonic phase\cite{pisarski2007}, inhomogeneous chiral condensed phase \cite{tatsumi2005,nickel2009,buballa2015}, etc.

Heavy ion collision experiments e.g. relativistic heavy ion collider (RHIC) and large hadron collider (LHC), give us
a unique opportunity to explore the QCD phase diagram. Strongly interacting QGP produced in these experiments at relativistic
energies recreates the physical conditions of the microsecond old universe just after the big bang. The strongly interacting plasma produced
in these high energy collisions can be characterized as high temperature and low baryon chemical potential QGP.
At high densities relative
to nuclear saturation density and low temperature exotic phases of QCD can exist, e.g. two flavor color superconducting
phase (2SC), color-flavor locked phase (CFL), crystalline color superconductor etc. Some of these high density QCD phases
can also be explored in the upcoming heavy ion collision experiments at moderate center of mass energies at FAIR and NICA.
Apart from these terrestrial experiments, interior of compact objects
like neutron stars give an ideal condition to
indirectly explore these high density QCD phases. Due to very low temperature and high baryon density, in the interior of a 
neutron star various QCD phases may be realized, e.g. meson condensation in hadronic phase\cite{kunihiro1993},
two flavor color superconducting phase, color-flavor locked phase \cite{Rajagopal2008,Rajagopal1999,byam2001} etc.

In the context of cold dense matter, compact objects like neutron stars can be strongly magnetized.  Observations 
indicates that the magnetic field strength at the surface of pulsars can be of the order of $10^{12}-10^{13}$ Gauss \cite{pulsarmag}.
Further, strongly magnetized neutron stars (magnetars) may have even stronger magnetic fields $\sim$ $10^{15}-10^{16}$ Gauss 
\cite{magnetar1,magnetar2,magnetar3,magnetar4,magnetar5,magnetar6,magnetar7}. Using virial theorem and comparing the magnetic field energy and
gravitational energy, one can estimate the physical upper bound on the strength of the magnetic field for a gravitationally bound star to 
be of the order $10^{18}$ Gauss\cite{pulsarmag}. For self bound objects like quark stars this bound can be even higher \cite{quarkstar}.
The physical origin of the very strong magnetic field in the magnetars
require reconsideration of the common understanding that the magnetic field of a neutron star is originated from the
progenitor stars \cite{Chanmugam1992}. Since quark matter can possibly be present at high densities, inside the neutron stars,
presence of quark ferromagnetic phase in high density quark matter has also been suggested as possible explanation of large 
magnetic field associated with magnetars
\cite{Tatsumi2000,Tatsumi2001,Tatsumi2003}. As a possible solution to this problem, author in Ref.\cite{Tatsumi2000} examined
the possible existence of spin-polarized deconfined quark matter using one gluon exchange interaction between quarks
in Fermi liquid theory within Hartree-Fock approximation. Taking the idea as proposed in the Ref.\cite{Tatsumi2000}, spin
polarization in the quark matter has been well explored in the subsequent literature. In general, a collective spin polarization of charged quarks
can give rise to ferromagnetic nature of quark matter at high density, hence the spin of the fermions play the crucial
role in determining the possibility of ferromagnetic nature of dense quark matter. It has been shown that in non-relativistic
framework there is no possibility of spin polarization in normal nuclear matter\cite{Pandharipande1972}. 
On the contrary, using relativistic Hartree-Fock approximation, possibility of spin polarization at asymptotic high density
has been suggested in Ref.\cite{Niembro1990,Niembro1991}. It is important to note that the relativistic framework may be
more suitable than the non-relativistic approach to understanding the existence of spin polarization. But in any case
to explore spin polarization in quark matter at a high density or baryon chemical potential relativistic approach is very natural.
In relativistic framework ``spin density'' can be expressed in two different ways, first by the spatial component of
the axial vector (AV) mean field, $\psi^{\dagger}\Sigma^i\psi\equiv-\bar{\psi}\gamma_5\gamma^i\psi$, constructed
out of the fermionic field (quarks)$\psi$ and axial vector combination of Dirac gamma matrices; second by tensor Dirac bilinear (T)
$\psi^{\dagger}\gamma^0\Sigma^i\psi\equiv-\bar{\psi}\sigma^{12}\psi$. Although AV and T type mean fields are different in the
massless limit of fermions, it has been shown that they are equivalent in nonrelativistic approximation \cite{Tatsumi2001}.
Coexistence of the spin polarization and color superconductivity has been studied using the AV interaction for quark matter in 
NJL model \cite{Tatsumi2003}. The interplay between the spin polarization and chiral symmetry breaking at finite density for
a single quark flavor using AV mean field has also been studied within NJL model in Ref.\cite{Maedan2007}. In Ref.\cite{Maedan2007},
it has been shown that for one flavor, spin polarization is possible at finite density and zero temperature provided
the ratio of the couplings of the axial vector channel and the pseudo scalar channel satisfies some lower bound. 
It has been
argued in Ref.\cite{Maedan2007} that due to the interplay between spin polarization and chiral symmetry for a certain value of chemical
potential, spin polarization appears due to the large dynamical quark masses generated by spontaneous chiral symmetry breaking.
Interestingly it
was also shown that spin polarization plays an important role in changing the value of the dynamical mass and at a very
high density, both dynamical quark mass and spin polarization vanish in the chiral symmetric phase. Although in
Ref.\cite{Tatsumi2000} author introduced the idea of quark spin polarization using one gluon exchange interaction,
in the NJL model studies, AV mean field has been used. 
Due to the Fierz transformation, one can get
AV channel interaction between quarks from one gluon exchange interaction, but the tensor Dirac bilinear
representation of ``spin density'' operator does not appear in the Fierz transformation of the one gluon exchange interaction.
Hence at asymptotically high densities where one gluon exchange interaction in perturbative QCD is applicable, spin
polarization cannot be studied using the T channel interaction. But for moderate densities near chiral phase transition
density perturbative QCD is not applicable and one can use QCD inspired low energy effective models e.g. NJL model. NJL model is not directly related
to perturbative one gluon exchange interaction. In this model AV or T, channel interaction is not written keeping in mind
the perturbative nature of QCD and some nonperturbative effects can give rise to tensor channel interaction.
Hence spin polarization in the tensor channel, which can be different from the AV channel can be studied within the NJL model.
In fact, the tensor channel opens up a completely different point of view in looking into the spin polarization
problem of quark matter at moderate densities e.g. spin-polarized phase can be shown to be present in the
chiral restored phase where the dynamical quark mass is zero \cite{Tsue2012,Tatsumi2011}. This result is different than the 
result obtained in Ref.\cite{Maedan2007}, where spin polarization is not present in chiral restored phase. Since the manifestation 
of the AV and T channel interaction is different, the interplay between the AV and T type spin-polarized phases becomes
interesting to study along with the other phases expected to arise in high baryon density region of the QCD phase
diagram \cite{Tatsumi2003,Tsue2013,Tsue2015,Tsue2017,Maedan2007,Tsue2012,Tsue2016,Tsue2018,tatsumi2005}.

In the present work we discuss the interplay between the spin polarization condensate ($\langle\bar{\psi}\Sigma^i\psi\rangle$) and 
the scalar chiral condensate $\langle\bar{\psi}\psi\rangle$ in (2+1) flavor NJL model using 
only tensor(T) type interaction for spin polarization. Most of the earlier works used some simplified approximation to study 
the interplay between spin polarization and other high density phases, which includes single flavor NJL model \cite{Maedan2007},
SU(2) flavor NJL model \cite{Tsue2012,Tsue2016}, 
SU(3) flavor NJL model \cite{Panda2013} with zero current quark mass etc.
However, for  a more realistic situation one should consider (2+1) flavor NJL
model with different current quark mass of strange and non-strange quarks. This apart, the structure of 
ferromagnetic condensation for (2+1) flavor NJL model is qualitatively different from that of two flavor NJL model as inherently 
two different kinds of spin polarizations are possible which are associated with the diagonal generators of the SU(3) flavour group. 
Behaviour of these spin polarization condensates as function of temperature and quark chemical potential $(\mu)$ has been discussed extensively.
Since the spin polarization condensates are also related to the quark-antiquark scalar condensates, it is evident that the spin polarization
condensates affect the constituent mass of the quarks. For a sufficiently large value of the tensor coupling the 
spin polarizations can also play an
important role in chiral phase transition, as will be shown, it can behave like a catalyst for the chiral symmetry restoration. In this work
spin polarization condensates due to the tensor type interaction appear in the chiral symmetry restored phase and the quark 
masses, specifically strange quark masses, are strongly affected by the spin polarization condensates in the chiral symmetric phase.

This paper is organized in the following manner. We first discuss the formalism of 2+1 flavour NJL model in the presence of tensor 
type interactions in Sec.\eqref{formalism}. In Sec.\eqref{formalism}  derivation of the thermodynamic potential is 
discussed in a mean field approach. Once the thermodynamic potential is derived  one can get the gap equations to solve for the 
condensates. After the formalism important results and the corresponding discussion are given in Sec.\eqref{results}. Finally in 
Sec.\eqref{conclusion} we summarize our work.

\section{formalism}
\label{formalism}
In order to study the spin polarization due to tensor channel interaction for realistic $(2+1)$ flavor and $SU(3)$ color quarks 
we start with the following NJL Lagrangian density \cite{Tsue2017,Buballa2005},

\begin{align}
 \mathcal{L}=\bar{\psi}\left(i\slashed{\partial}-\hat{m}\right)\psi+\mathcal{L}_{sym}+\mathcal{L}_{det}+\mathcal{L}_{tensor}
 +\mu\bar{\psi}\gamma^0\psi,
 \label{equn1}
\end{align}
where $\psi=(u , d, s)^T$ is the three flavor quark field and the diagonal current quark matrix is
$\hat{m}=\text{diag}_f(m_u,m_d,m_s)$. In this work we have assumed that due to isospin symmetry in the non strange quark sector
$m_u=m_d$. Strange quark mass $m_s$ is different from the other light quark masses. Difference between the strange and non strange 
quark masses explicitly breaks the $SU(3)$ flavor symmetry. $\mu$ is the quark chemical potential. In literature different chemical
potential for the strange and nonstrange quarks have been considered, but the phase diagram has no qualitative difference.
In this case we are assuming that the quark chemical potential of the strange and nonstrange quarks are same.
Following the representations of different interaction terms as given in Ref.\cite{Buballa2005}, in general one considers, 

\begin{align}
 \mathcal{L}_{sym} = g\sum_{a=0}^{a=8}\bigg[\left(\bar{\psi}\lambda_a\psi\right)^2
 +\left(\bar{\psi}i\gamma_5\lambda_a\psi\right)^2\bigg].
 \label{equn2}
\end{align}
This term has been constructed keeping in mind the $U(3)_L\times U(3)_{R}$ chiral symmetry for three flavor case and it can be 
generalized to any number of flavours $N_f$. 
The interaction term $\mathcal{L}_{sym}$ represents four point interaction, where $\lambda_0=\sqrt{2/3}I_f$ and $\lambda_a$,
$a=1,....(N_f^2-1)$ are the generators of $SU(N_f)$. In the present case $I_f$ is $3\times3$ identity matrix and 
$\lambda_a$ for $a=1,...8$ are the Gell-Mann matrices.  

The interaction term $\mathcal{L}_{det}$ in Eqn.\eqref{equn1} is `t Hooft determinant interaction term. This term breaks $U(1)$ axial symmetry 
explicitly in QCD and also successfully describes the nonet meson properties\cite{Weise1991,Hatsuda1994,Klevansky1996}. It
can be expressed as, 

\begin{align}
\mathcal{L}_{det}= -K \text{det}_f[\bar{\psi}(1+\gamma_5)\psi+h.c]
\label{equn4}
\end{align}
In this interaction term determinant is taken in the flavour space. This term represents maximally flavour-mixing $2N_f$ point 
interaction for $N_f$ quark flavours. For two flavour NJL model this term does not introduce any new dynamics because 
for two flavour case it gives four Fermi interaction which is already there. But for three or 
more flavours this term generates new type of interaction, e.g. for three flavour case it gives rise to six point interaction term.
The tensor interaction which is responsible for spin polarization is given as \cite{Tsue2017, Panda2013},

\begin{align}
 \mathcal{L}_{tensor}=\frac{G_T}{2}\sum_{a=3,8}\left(\bar{\psi}\Sigma_z\lambda_a\psi\right)^2,
 \qquad \Sigma_z= \left(\begin{array}{cc} \sigma_z & 0\\ 0 & \sigma_z \end{array}\right),
 \label{equn5}
\end{align}
 where $\sigma_z$ is the third Pauli matrix. Here we have assumed polarization along the z-axis. Note that $\mathcal{L}_{tensor}$ is not invariant under chiral symmetry, rather 
one requires to add a similar term with $\gamma^5$ matrix to make the tensor interaction symmetric under chiral symmetry. Since 
we are not considering any condensation involving $\gamma^5$, we have omitted the term which ensures chiral invariance for the tensor 
interaction. Thus the total Lagrangian with finite chemical potential becomes,

\begin{align}
\mathcal{L}=\bar{\psi}\left(i\slashed{\partial}-\hat{m}\right)\psi
+g\sum_{a=0}^{a=8}\left(\bar{\psi}\lambda_a\psi\right)^2-K \text{det}_f[\bar{\psi}(1+\gamma_5)\psi+h.c]
+\sum_{a=3,8}\frac{G_T}{2}\left(\bar{\psi}\Sigma_z\lambda_a\psi\right)^2+\mu\bar{\psi}\gamma^0\psi.
\label{equn6}
\end{align}

In mean field approximation expanding the operators around their expectation values and neglecting higher order fluctuations,
we obtain,

\begin{align}
 &\left(\bar{u}u\right)^2\simeq2\langle\bar{u}u\rangle\bar{u}u-\langle\bar{u}u\rangle^2=2\sigma_{ud}\bar{u}u-\sigma_{ud}^2\nonumber\\
 &\left(\bar{d}d\right)^2\simeq2\langle\bar{d}d\rangle\bar{d}d-\langle\bar{d}d\rangle^2=2\sigma_{ud}\bar{d}d-\sigma_{ud}^2\nonumber\\
 &\left(\bar{s}s\right)^2\simeq2\langle\bar{s}s\rangle\bar{s}s-\langle\bar{s}s\rangle^2=2\sigma_{s}\bar{s}s-\sigma_{s}^2\nonumber\\
 &\left(\bar{\psi}\Sigma_z\lambda_3\psi\right)^2\simeq 2\langle\bar{\psi}\Sigma_z\lambda_3\psi\rangle
 \left(\bar{\psi}\Sigma_z\lambda_3\psi\right)-\langle\bar{\psi}\Sigma_z\lambda_3\psi\rangle^2
 =2F_3\left(\bar{\psi}\Sigma_z\lambda_3\psi\right)-F_3^2\nonumber\\
 &\left(\bar{\psi}\Sigma_z\lambda_8\psi\right)^2\simeq 2\langle\bar{\psi}\Sigma_z\lambda_8\psi\rangle
 \left(\bar{\psi}\Sigma_z\lambda_8\psi\right)-\langle\bar{\psi}\Sigma_z\lambda_8\psi\rangle^2
 =2F_8\left(\bar{\psi}\Sigma_z\lambda_8\psi\right)-F_8^2,
 \label{equn7}
\end{align}

where the chiral condensates or the quark-antiquark condensates are $\langle\bar{u}u\rangle=\langle\bar{d}d\rangle\equiv\sigma_{ud}$,
$\langle\bar{s}s\rangle \equiv \sigma_{s}$ and the spin polarization condensates are
$F_3=\langle\bar{\psi}\Sigma_z\lambda_3\psi\rangle$ and $F_8=\langle\bar{\psi}\Sigma_z\lambda_8\psi\rangle$. We can write the 
mean field Lagrangian as,

\begin{align}
 \mathcal{L}=\bar{\psi}\left(i\slashed{\partial}-\hat{M}+G_TF_3\Sigma_z\lambda_3+G_TF_8\Sigma_z\lambda_8+\mu\gamma^0\right)\psi
 &-2g\left(\sigma_{ud}^2+\sigma_{ud}^2+\sigma_{s}^2\right)+4K\sigma_{ud}^2\sigma_{s}\nonumber\\
 &-\frac{G_T}{2}F_3^2-\frac{G_T}{2}F_8^2,
 \label{equn8}
 \end{align}

where, $\hat{M}\equiv\text{diag}(M_u,M_d,M_s)$, with effective masses,

\begin{align}
M_u=&m_u-4g\sigma_{ud}+2K\sigma_{ud}\sigma_s\nonumber\\
M_d=&m_d-4g\sigma_{ud}+2K\sigma_{ud}\sigma_s\nonumber\\
M_s=&m_s-4g\sigma_{s}+2K\sigma_{ud}^2.
\end{align}

For a given system at finite temperature and finite chemical potential most important quantity for the understanding of the thermodynamic
behaviour or the phase structure, is the thermodynamic potential. Once the thermodynamic potential for this model is known,
thermodynamic quantities can be extracted using Maxwell relations. The thermodynamic potential for the Lagrangian as given in 
Eqn.\eqref{equn8} in the grand canonical ensemble at 
a finite temperature and finite chemical potential can be given as:
 
 \begin{align}
\Omega(T,\mu,\sigma_{ud},\sigma_s,F_3,F_8)=&-N_c\sum_{f=u,d,s}\int\frac{d^3p}{(2\pi)^3}\bigg[\bigg(E_{f+}+E_{f-}\bigg)
+T\ln\bigg(1+e^{-\beta(E_{f+}-\mu)}\bigg)\nonumber\\
&+T\ln\bigg(1+e^{-\beta(E_{f+}+\mu)}\bigg)
+T\ln\bigg(1+e^{-\beta(E_{f-}-\mu)}\bigg)\nonumber\\
&+T\ln\bigg(1+e^{-\beta(E_{f-}+\mu)}\bigg)\bigg]\nonumber\\
& +2g(\sigma_{ud}^2+\sigma_{ud}^2+\sigma_s^2)-4K\sigma_{ud}^2\sigma_s+\frac{G_T}{2}F_3^2+\frac{G_T}{2}F_8^2,\nonumber\\
= & -\frac{6}{4\pi^2}\int_0^{\Lambda}dp_T\int_0^{\sqrt{\Lambda^2-p_T^2}}p_T dp_z\bigg[\bigg(E_{f+}+E_{f-}\bigg)
+T\ln\bigg(1+e^{-\beta(E_{f+}-\mu)}\bigg)\nonumber\\
&+T\ln\bigg(1+e^{-\beta(E_{f+}+\mu)}\bigg)
+T\ln\bigg(1+e^{-\beta(E_{f-}-\mu)}\bigg)\nonumber\\
&+T\ln\bigg(1+e^{-\beta(E_{f-}+\mu)}\bigg)\bigg]\nonumber\\
& +2g(\sigma_{ud}^2+\sigma_{ud}^2+\sigma_s^2)-4K\sigma_{ud}^2\sigma_s+\frac{G_T}{2}F_3^2+\frac{G_T}{2}F_8^2
\label{eq2}
\end{align}
where $N_c=3$ is the number of colors, transverse momentum $p_T=\sqrt{p_x^2+p_y^2}$ and the single particle energies are, 

\begin{align}
E_{u+}=&\sqrt{p_z^2+\bigg(\sqrt{p_T^2+M_u^2}+G_T\bigg(F_3+\frac{F_8}{\sqrt{3}}\bigg)\bigg)^2}\nonumber\\
E_{u-}=&\sqrt{p_z^2+\bigg(\sqrt{p_T^2+M_u^2}-G_T\bigg(F_3+\frac{F_8}{\sqrt{3}}\bigg)\bigg)^2}\nonumber\\
E_{d+}=&\sqrt{p_z^2+\bigg(\sqrt{p_T^2+M_d^2}+G_T\bigg(F_3-\frac{F_8}{\sqrt{3}}\bigg)\bigg)^2}\nonumber\\
E_{d-}=&\sqrt{p_z^2+\bigg(\sqrt{p_T^2+M_d^2}-G_T\bigg(F_3-\frac{F_8}{\sqrt{3}}\bigg)\bigg)^2}\nonumber\\
E_{s+}=&\sqrt{p_z^2+\bigg(\sqrt{p_T^2+M_s^2}+G_T\frac{2F_8}{\sqrt{3}}\bigg)^2}\nonumber\\
E_{s-}=&\sqrt{p_z^2+\bigg(\sqrt{p_T^2+M_s^2}-G_T\frac{2F_8}{\sqrt{3}}\bigg)^2}\nonumber\\
\label{dispersion}
\end{align}

Thermodynamic behaviour of the condensates can be found by solving the gap equations, which can be found from the stationary conditions.

\begin{align}
 \frac{\partial\Omega}{\partial\sigma_{ud}}=\quad  \frac{\partial\Omega}{\partial\sigma_{s}}= \quad 
 \frac{\partial\Omega}{\partial F_3}= \quad  \frac{\partial\Omega}{\partial F_8}=0
 \label{gapeq}
\end{align}
Gap equations can have several roots, but the solution with the lowest value of thermodynamic potential is taken as the stable 
solution.

NJL model Lagrangian in (3+1) dimension has operators which have mass dimension more than four, thus it can shown to be a 
non-renormalizable theory \cite{Klevansky1992}. Thus the divergence coming from the three momentum integral of the vacuum part
can not be removed by the renormalization prescriptions. The model predictions inevitably depend on the regularization procedures
and parameter dependence in each regularization method has been reported in Ref.\cite{Kohyama2016,Kohyama2015}. 
In this work we have considered the  most frequently used 3D momentum cutoff regulation scheme to regularize the divergence in Eq.\eqref{eq2}
for thermodynamic potential.

\begin{figure}[] 
 \begin{center}
  \includegraphics[width=0.6\linewidth]{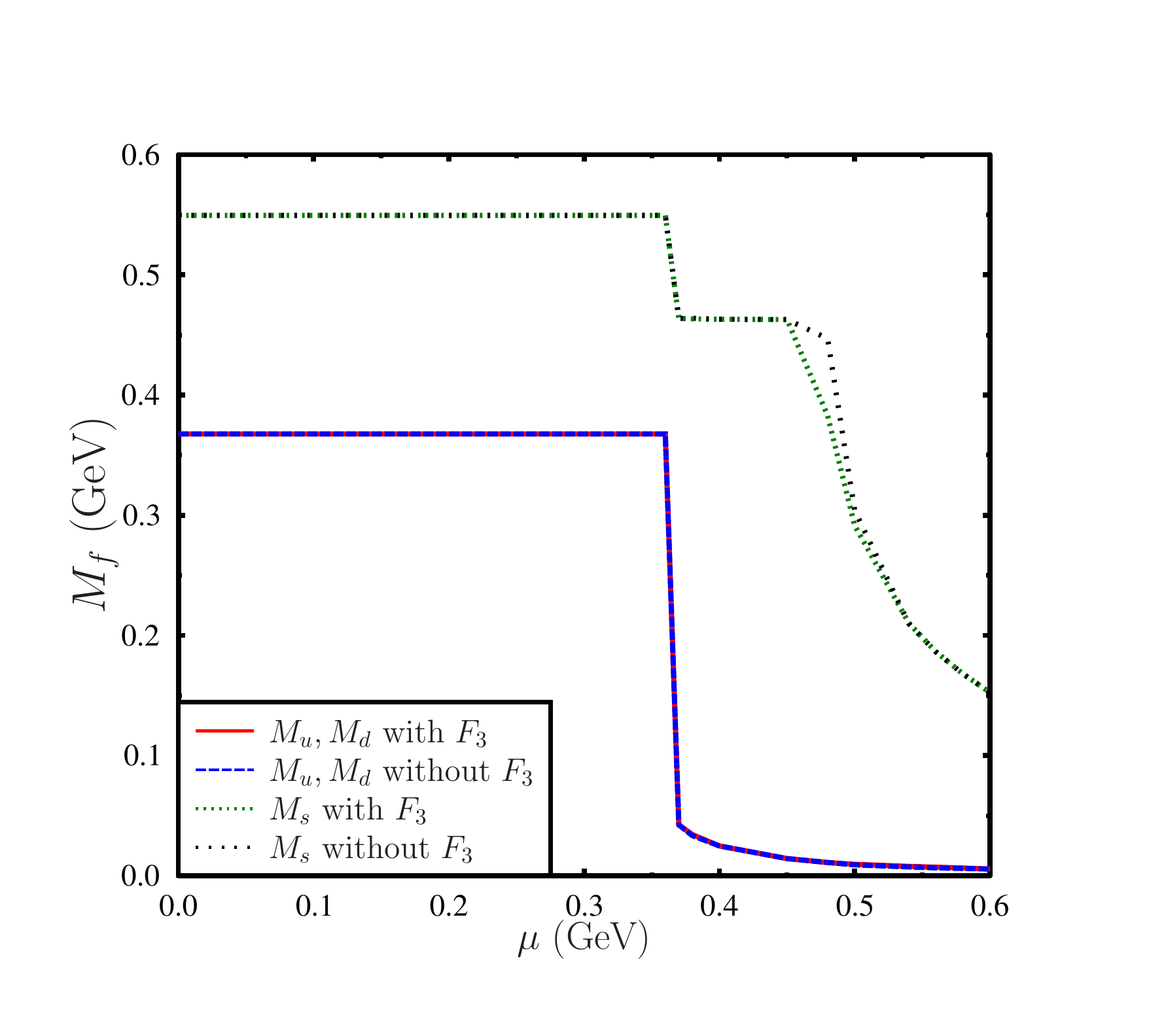} 
 \end{center}
 \caption{Constituent quark mass as a function of quark chemical potential at zero temperature in the presence and absence of 
 spin polarization condensation. Blue-dashed line and green-dotted line represent non strange and strange quark mass in the
 presence of spin polarization condensate $F_3$. Red-solid line and black-dotted line represents non strange and strange quark
 constituent mass in the standard 2+1 flavor NJL model in the absence of any spin polarization condensate. Sharp jump in the 
 value of $M_u$ and $M_s$ near $\mu=0.360$ GeV indicates the first order chiral phase transition. In this case we have considered
 the tensor interaction coupling to be $G_T=2g$. Comparing green and the blue 
 lines for strange quark it is clear that non zero value of spin condensate affects strange quark mass. However, the non strange
 quark masses are almost unaffected due to the presence of spin polarization condensate. For $G_T=2g$ non zero value of  $F_3$ 
 appears only near 0.480 GeV which is away from the chiral phase transition critical chemical potential, hence in this 
 case the chiral phase transition is unaffected by the presence of spin polarization.}
 \label{massfig1}
\end{figure}

\begin{figure}[] 
    \begin{minipage}[b]{0.5\linewidth}
    \centering
    \includegraphics[width=0.8\linewidth]{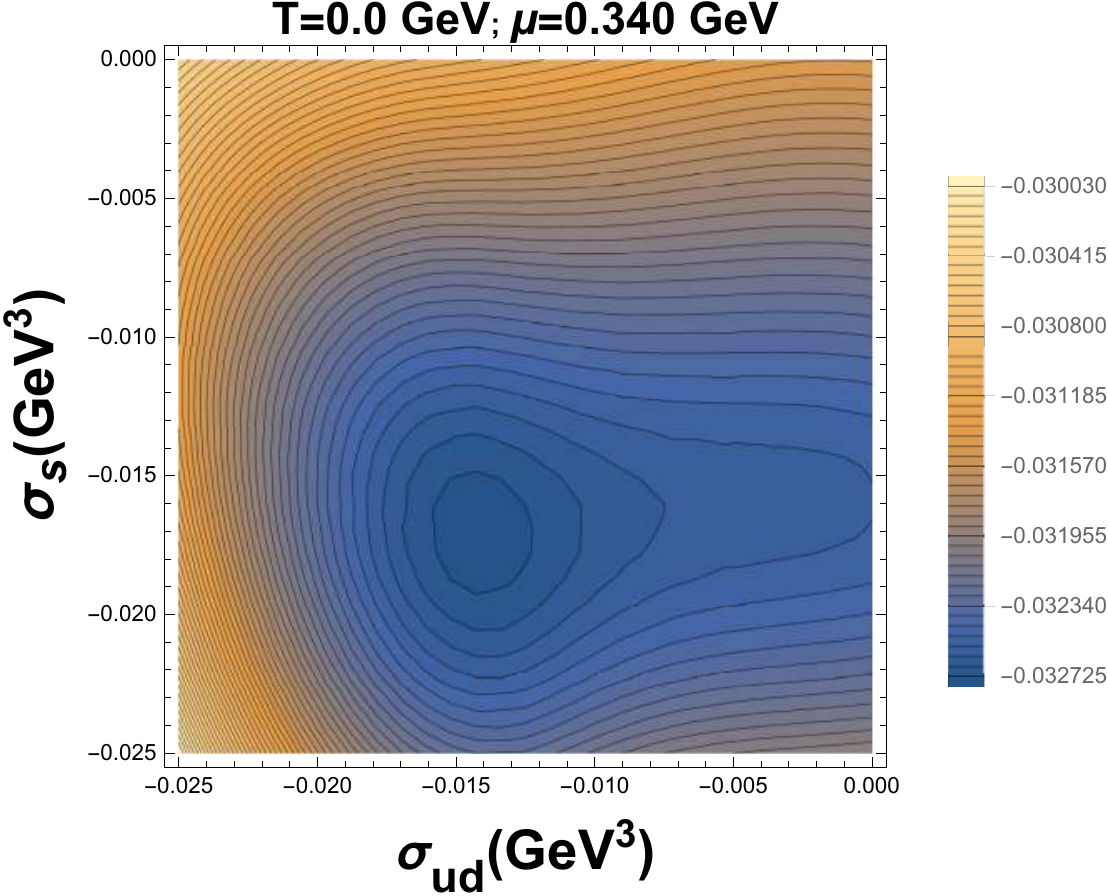} 
    \vspace{4ex}
  \end{minipage}
  \begin{minipage}[b]{0.5\linewidth}
    \centering
    \includegraphics[width=0.8\linewidth]{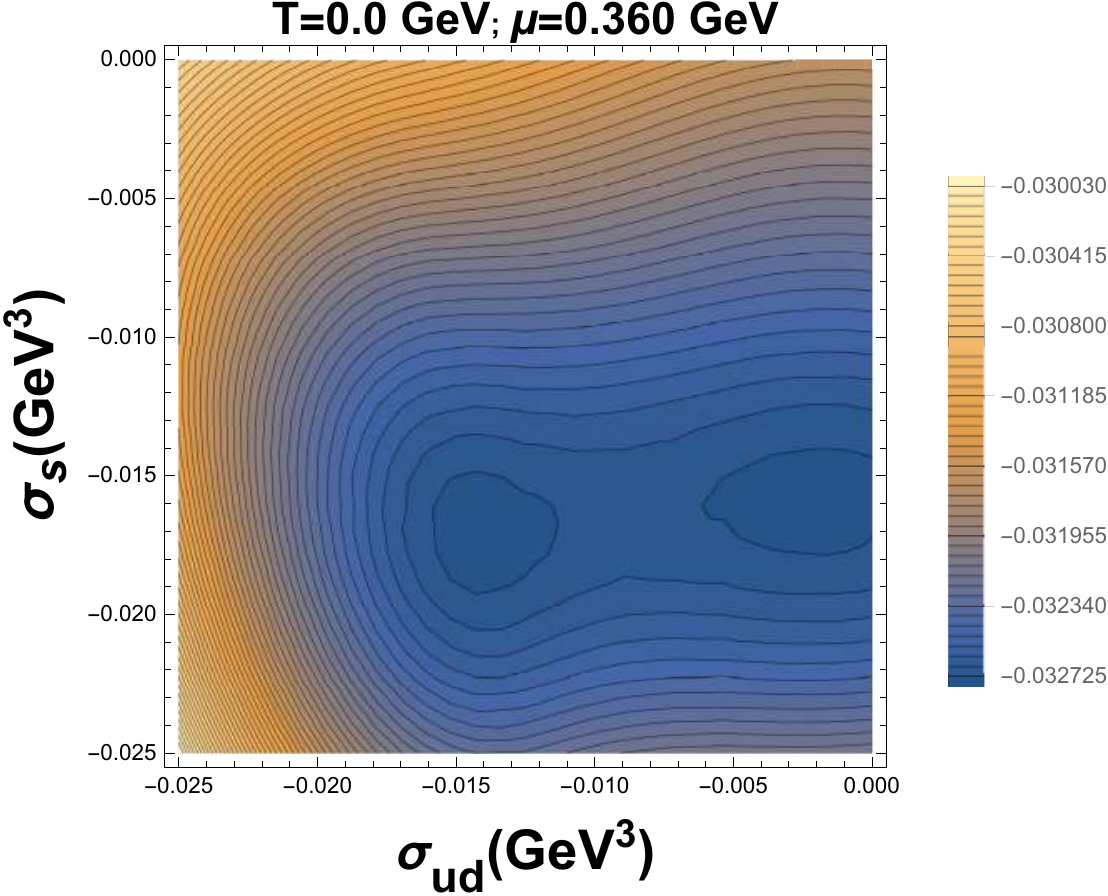} 
    \vspace{4ex}
  \end{minipage} 
  \begin{minipage}[b]{0.5\linewidth}
    \centering
    \includegraphics[width=0.8\linewidth]{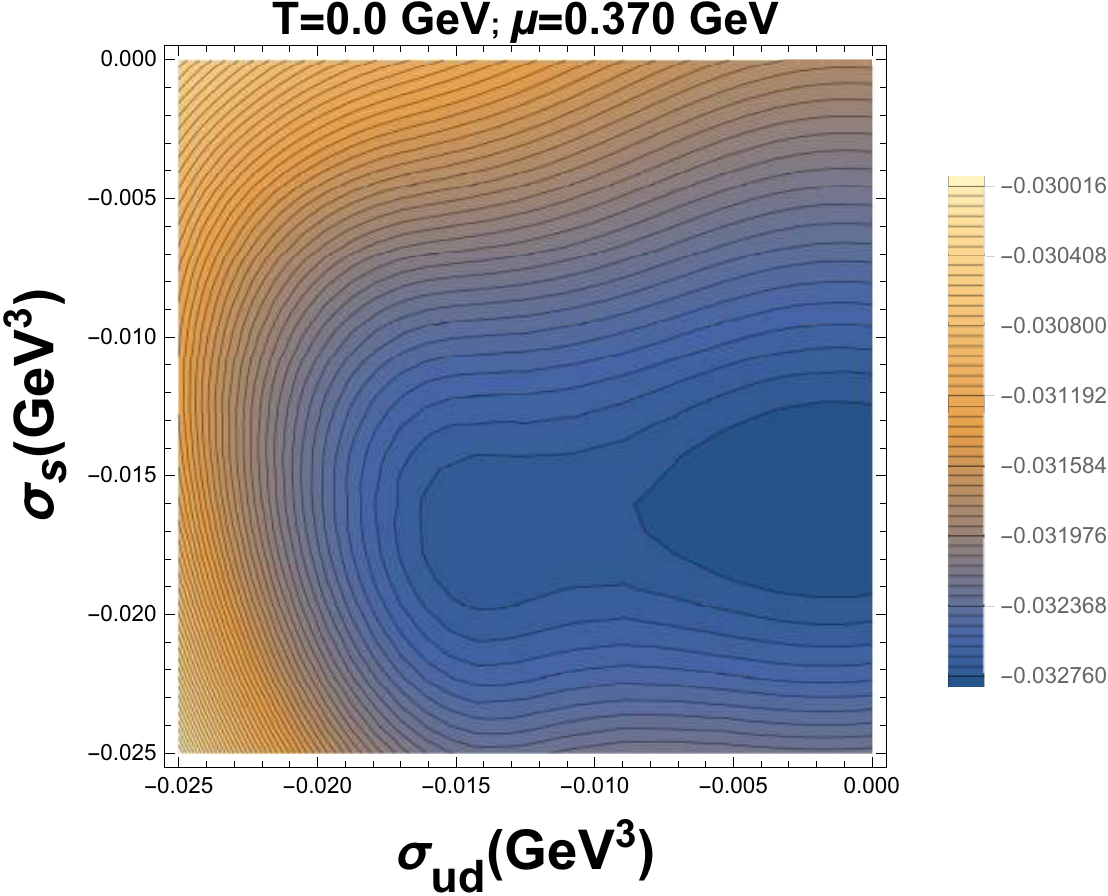} 
    \vspace{4ex}
  \end{minipage}
  \begin{minipage}[b]{0.5\linewidth}
    \centering
    \includegraphics[width=0.8\linewidth]{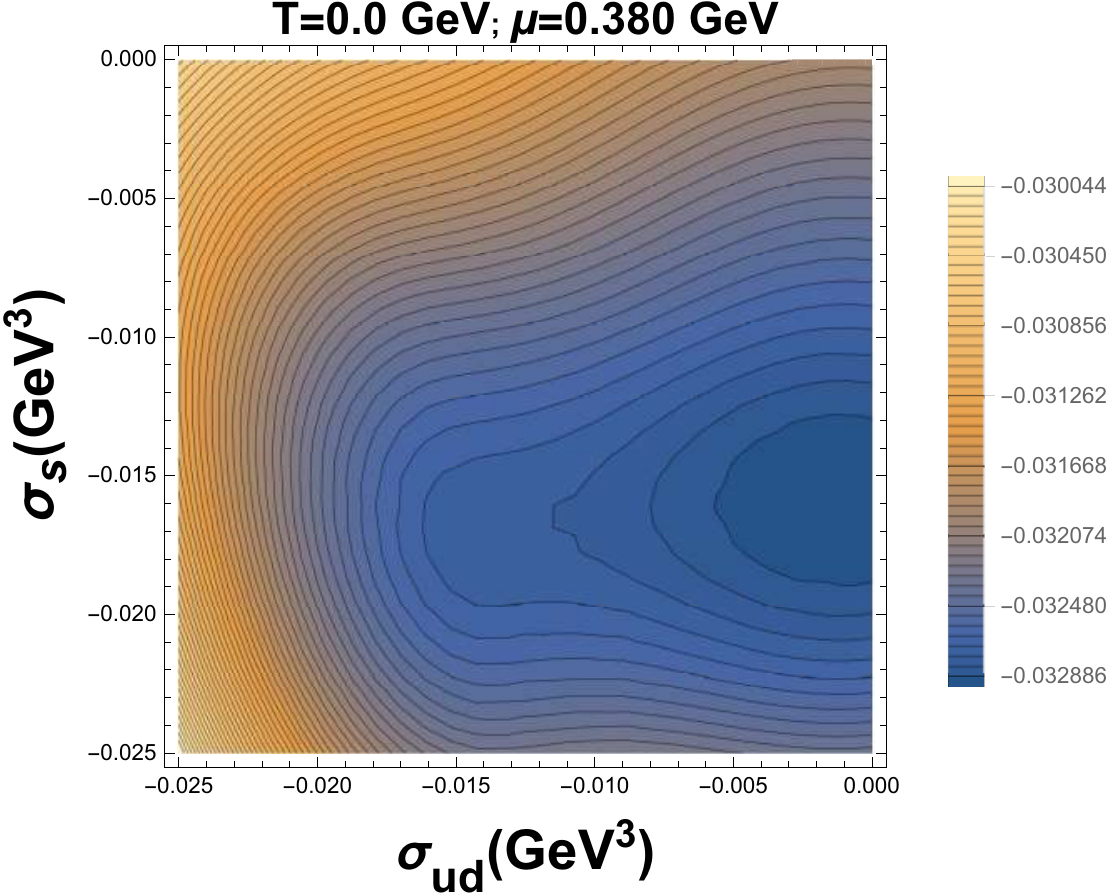} 
    \vspace{4ex}
  \end{minipage} 
  \caption{The figure shows the contour maps of the thermodynamics potential with the set of parameters in table\eqref{table1} and
  $G_T=2g$ at $T=0.0$ GeV for different values of $\mu$. The darker region in the plots show the lower value of the thermodynamic
  potential. The horizontal and vertical axes represents the non strange quark condensate $\sigma_{ud}$ and strange quark condensate
  $\sigma_s$ respectively. Existence of almost degenerate vacuum is clear from the figure near $\mu=0.360$ GeV. Hence the chiral 
  phase transition near $\mu=0.360$ is a first order phase transition. Spin polarization condensation $F_3$ has no effect on the chiral phase transition. As we have shown in Fig.\eqref{fig2}
  non zero value of $F_3$ occurs near 
  $\mu=0.480$ GeV at $T=0.0$GeV for $G_T=2g$, which is far away from the critical quark chemical potential for the 
  chiral phase transition.}
  \label{fig1}
\end{figure}

In the study of spin polarization in NJL model, the parameter which plays the crucial role is the tensor channel interaction $G_T$.
If one considers only vector current interaction, e.g. one gluon exchange interaction in perturbative QCD processes, then such a 
tensor interaction can not be generated by Fierz transformation. However, such a tensor interaction can be generated from 
two gluon exchange diagrams \cite{Tsue2017}. It is relevant to point out that one can also get tensor channel interaction by
Fierz transformation from scalar and pseudo scalar interaction \cite{Tsue2012},

\begin{align}
 g\bigg[(\bar{\psi}\psi)^2+(\bar{\psi}i\gamma_5\lambda_a\psi)^2\bigg]=\frac{g}{4}\bigg[(\bar{\psi}\psi)^2
 -\frac{1}{2}(\bar{\psi}\gamma^{\mu}\gamma^{\nu}\lambda_a\psi)^2+.....\bigg],
\end{align}
which gives $|g/G_T|=2$. In the present investigation we can take $G_T$ as a free parameter to study the inter relationship between
scalar and tensor condensates. It may also be noted that the parameters $g$ and $G_T$ may be considered independently to 
derive mesonic properties \cite{Jaminon1998,Jaminon2002,Battistel2016}. It has been shown that $SU(2)$ NJL model with both positive 
and negative tensor couplings can describe the phenomenology of mesons. Indeed $SU(2)$ Lagrangian has been considered with 
vector, axial vector and tensor interaction in Ref. \cite{Battistel2016} where, the gap equations are solved in the 
in the usual Hartree approximation while mesons are described in the random phase approximation \cite{Battistel2016}. In this work 
we have only considered $G_T$ as a free parameter with positive values only i.e. $G_T$ and $g$ are of same sign. In the literature
various values have been considered e.g.  $G_T=2g, 1.5g$ \cite{Tsue2017} as well as $G_T=4.0 g$ \cite{Battistel2016}. We have also
obtained our results taking different values of $G_T$. Results with some specific parameter sets have been mentioned in the result 
and discussion section.

\begin{figure}[!h] 
    \begin{minipage}[b]{0.5\linewidth}
    \centering
    \includegraphics[width=0.95\linewidth]{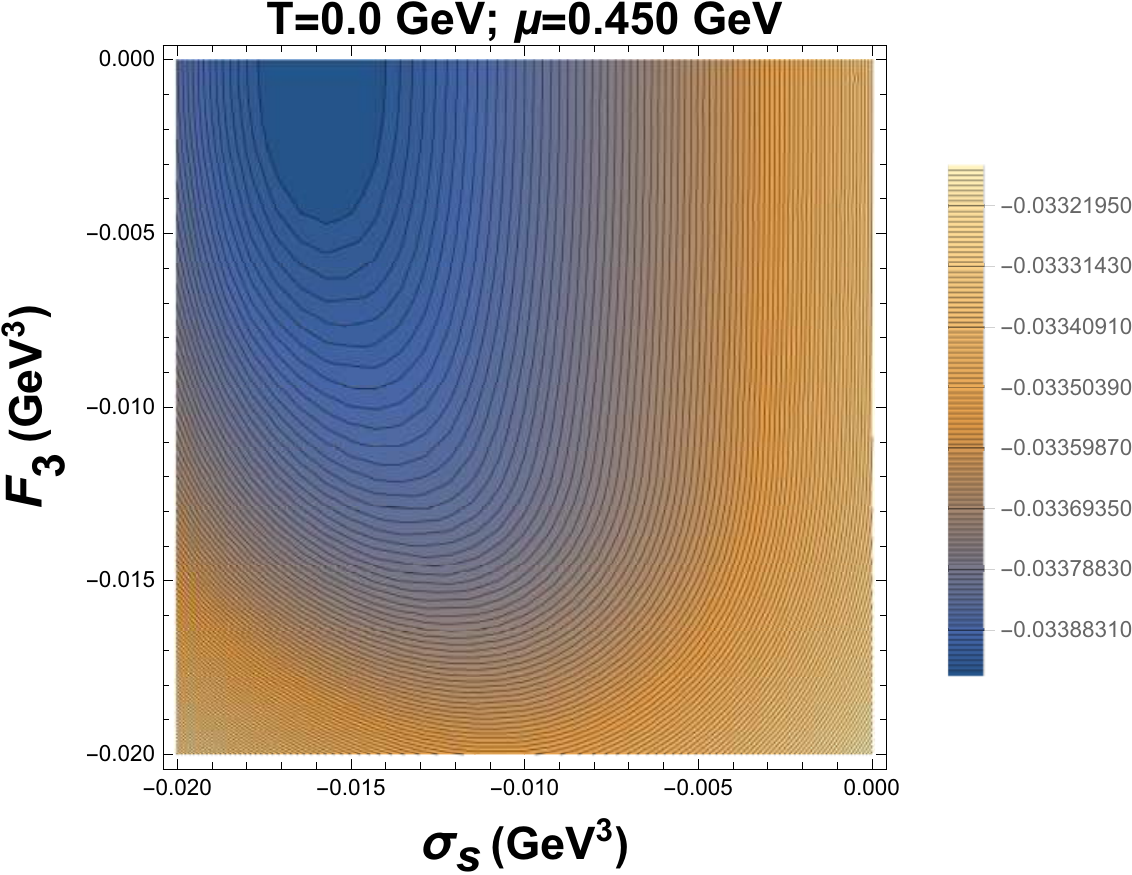} 
    \vspace{4ex}
  \end{minipage}
  \begin{minipage}[b]{0.5\linewidth}
    \centering
    \includegraphics[width=0.95\linewidth]{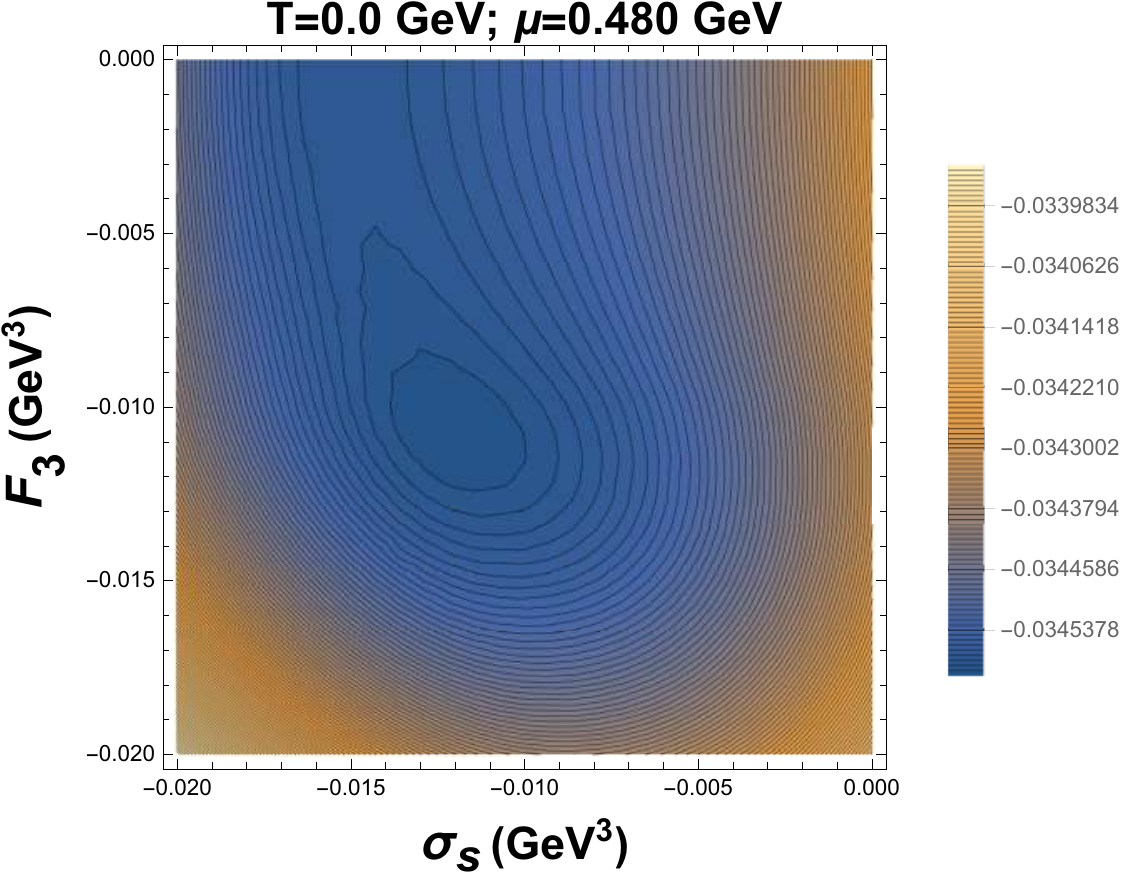} 
    \vspace{4ex}
  \end{minipage} 
  \begin{minipage}[b]{0.5\linewidth}
    \centering
    \includegraphics[width=0.95\linewidth]{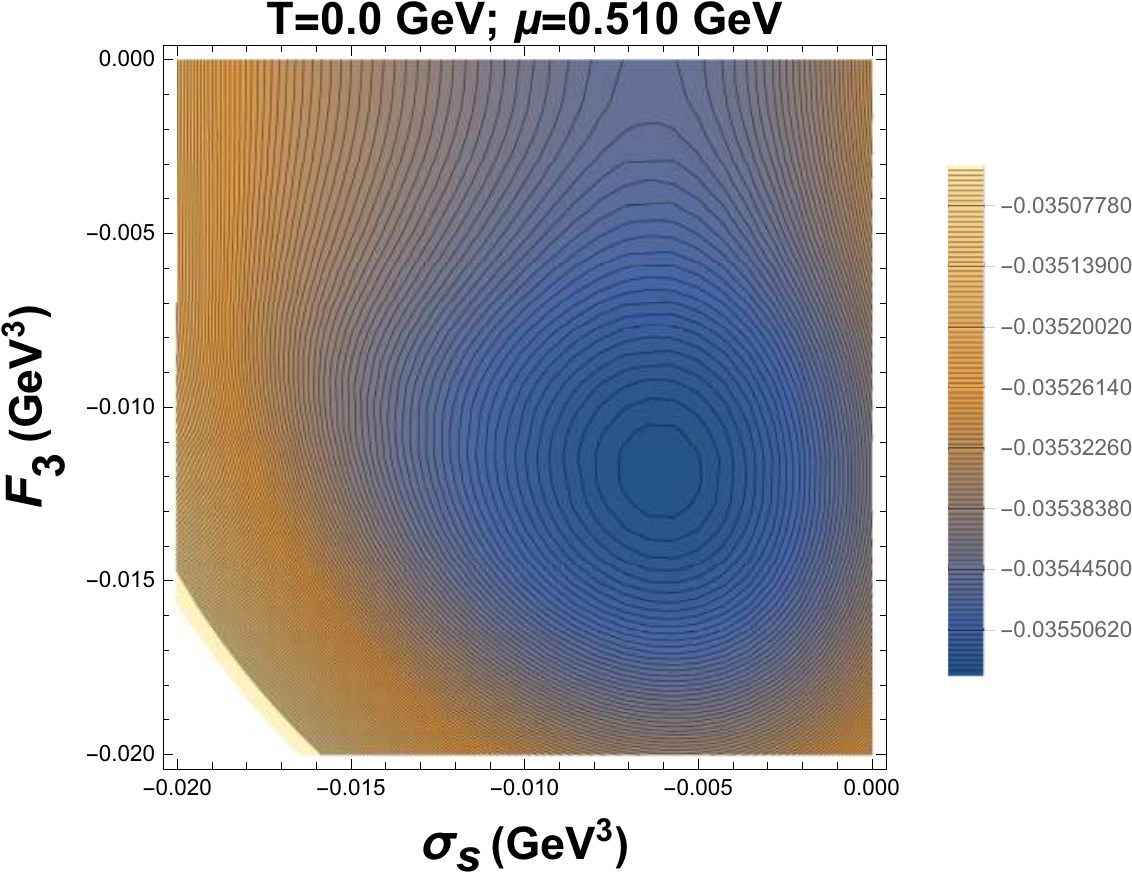} 
    \vspace{4ex}
  \end{minipage}
  \begin{minipage}[b]{0.5\linewidth}
    \centering
    \includegraphics[width=0.95\linewidth]{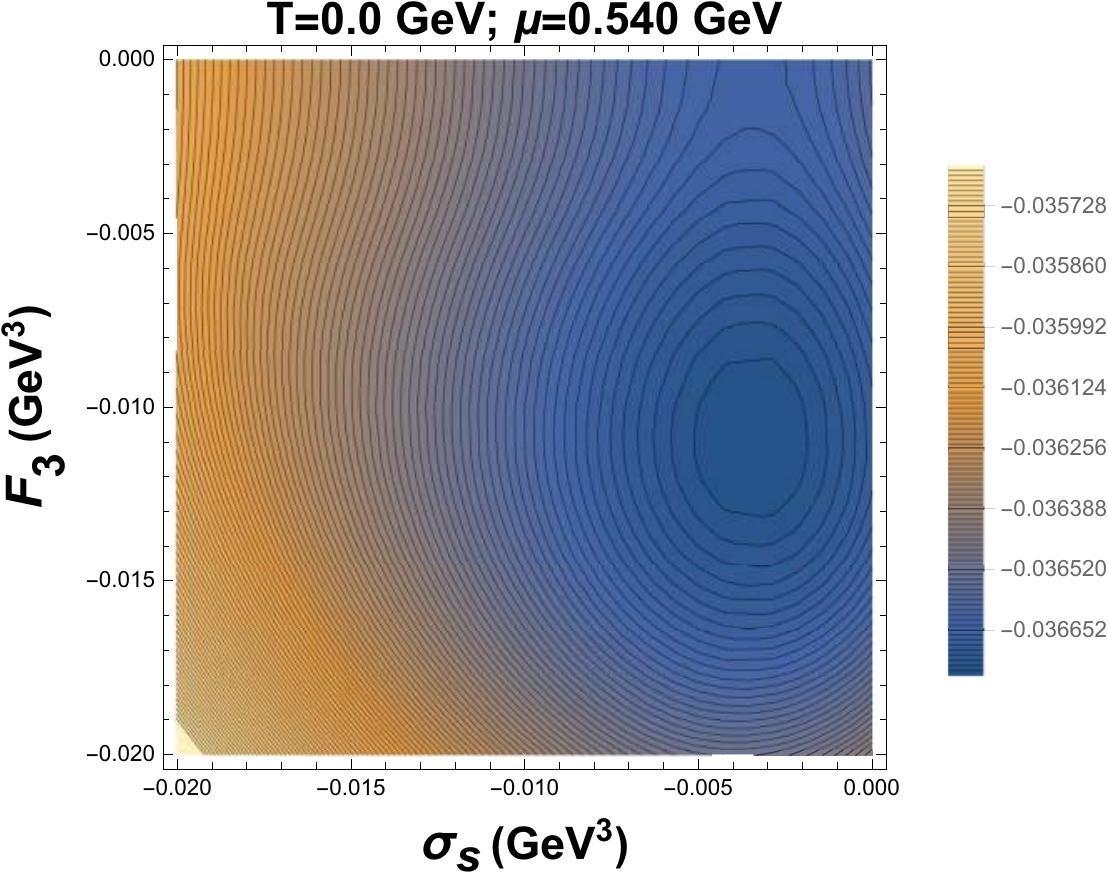} 
    \vspace{4ex}
  \end{minipage}
  \begin{minipage}[b]{0.5\linewidth}
    \centering
    \includegraphics[width=0.95\linewidth]{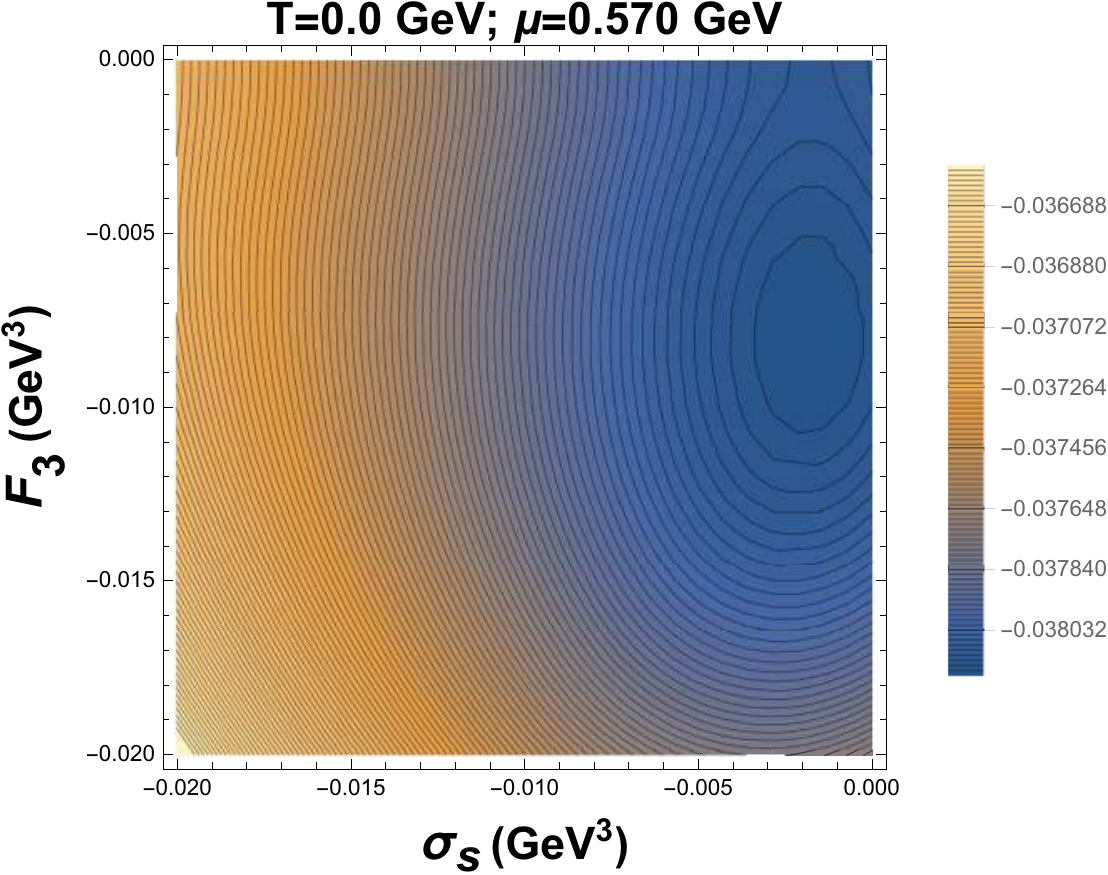} 
    \vspace{4ex}
  \end{minipage}
  \begin{minipage}[b]{0.5\linewidth}
    \centering
    \includegraphics[width=0.95\linewidth]{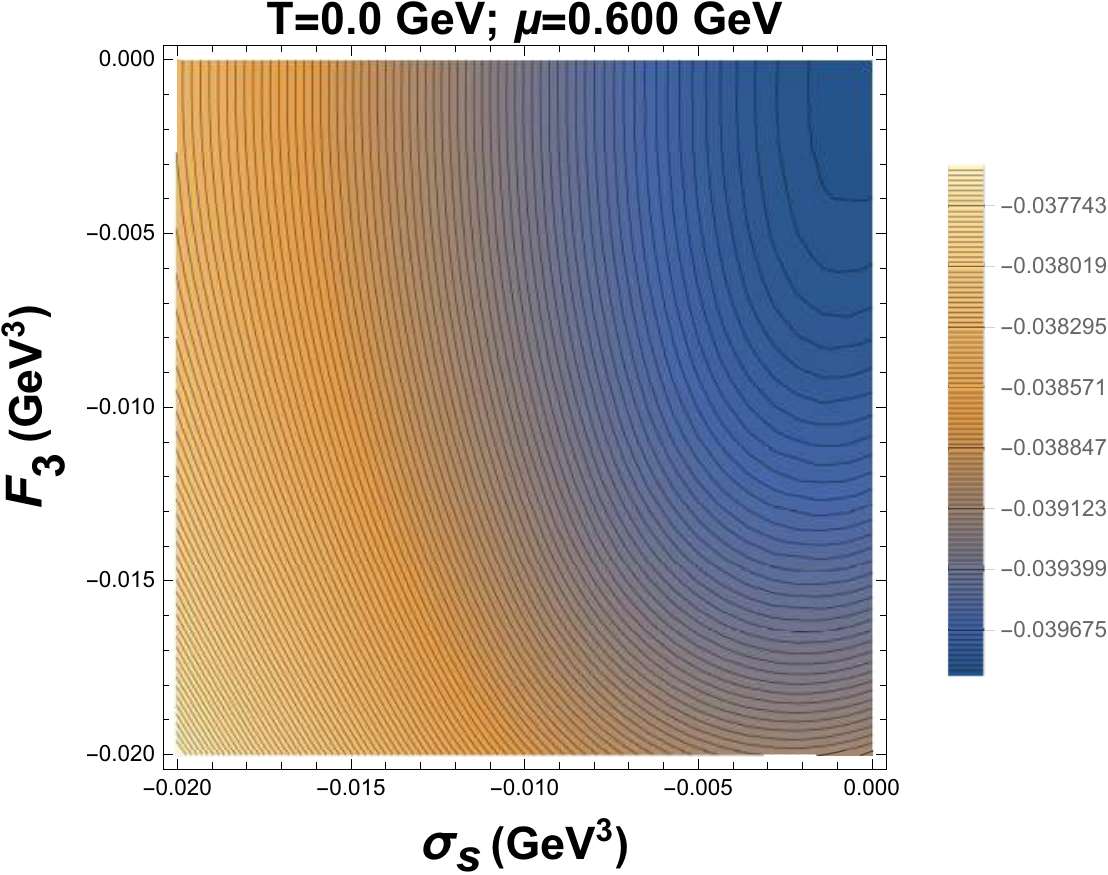} 
    \vspace{4ex}
  \end{minipage} 
  \caption{This figure shows the contour plots of the thermodynamic potential in $\sigma_s-F_3$ plane at zero temperature with 
  different values of quark chemical potentials ($\mu$) for the case of $G_T=2g$ and $F_8=F_3/\sqrt{3}$. It is clear from the plots that non zero spin polarization appears at $\mu=0.480$ GeV, reaches its maximum value near 
  $\mu=0.510$ GeV and it completely melts near $\mu=0.600$ GeV.}
  \label{fig2}
\end{figure}

\section{Results and Discussions}
\label{results}

We begin the discussion with the parameterization of the model. The parameters to be fixed are the three current quarks masses
($m_{u}, m_{d}, m_{s}$),
 the scalar coupling $(g)$, the determinant coupling $K$, the tensor coupling$(G_T)$ and the three momentum
 cut-off $\Lambda$ to regularize divergent integrals. Except 
 for the tensor coupling $G_T$, there are several parameter sets available for NJL model \cite{Buballa2005}. These fits are obtained using low energy 
 hadronic properties such as pion decay constant and masses of pion, kaon and $\eta^{\prime}$ \cite{Hatsuda1994,Lutz1992,Rehberg1996}.
 The determinant interaction is important as it 
 breaks $U(1)_A$ symmetry and gives correct $\eta$ mass. One may note that there is discrepancy in determination of the determinant
 coupling K. For example in Ref. \cite{Hatsuda1994} the value of the coupling differs by as much as 30 percent compared to value used in present work. This
 discrepancy arises due to difference in treatment of $\eta^{\prime}$ mesons with a high mass \cite{Buballa2005}. In fact, this leads to an nonphysical imaginary part for the corresponding polarization diagram in the $\eta^{\prime}$ meson channel.
This is unavoidable because NJL is not confining and is unrealistic in this context. Within the above mentioned limitations 
of the model and the uncertainty in the value of the determinant coupling, we proceed
with the present parameter set as given in Table \eqref{table1} \cite{Buballa2005}.

\begin{table}[h]
\begin{tabular}{ |p{5cm}||p{4cm}| }
 \hline
 \multicolumn{2}{|c|}{\textbf{Parameter Set}} \\
 \hline
  \textbf{Parameters and couplings}& \textbf{Value} \\
 \hline
 Three momentum cutoff ($\Lambda$)  & $\Lambda=602.3\times10^{-3}$ (GeV) \\
 \hline
 $u$ quark mass $(m_u)$ &  $m_u = 5.5 \times 10^{-3}$ (GeV) \\
 \hline
 $d$ quark mass $(m_d)$ &  $m_d = 5.5 \times 10^{-3}$ (GeV) \\
\hline
$s$ quark mass $(m_s)$ &  $m_s = 140.7 \times 10^{-3}$ (GeV) \\
\hline
Scalar coupling $(g)$ & $g=1.835/\Lambda^2$\\
\hline
Determinant interaction $(K)$ & $K=12.36/\Lambda^5$\\
\hline
\end{tabular}
\caption{Parameter set considered in this work for 2+1 NJL model apart from the tensor coupling $G_T$.}
\label{table1}
\end{table}

Let us first note that there are four condensates, $\sigma_{ud}$, $\sigma_s$,
$F_3\equiv \langle \bar{u} \Sigma_z u\rangle - \langle \bar{d} \Sigma_z d\rangle$ and
$F_8\equiv \frac{1}{\sqrt{3}}\left(\langle \bar{u} \Sigma_z u\rangle + \langle \bar{d} \Sigma_z d\rangle -
2 \langle \bar{s} \Sigma_z s\rangle\right)$, to be determined from the solution of the gap Eq.\eqref{gapeq}. However for simplicity
we shall first consider $F_8=\frac{F_3}{\sqrt{3}}$, so that the spin polarization condensate for $d$ quarks and $s$ quarks are treated
at the same footing i.e. $\langle \bar{d}\Sigma_z d\rangle\equiv \langle \bar{s}\Sigma_z s\rangle$ \cite{Panda2013}. The results in such a
scenario is determined below.

\subsection{Results with $F_8$=$\frac{F_3}{\sqrt{3}}$}
\subsubsection{Chiral phase transition and the behavior of quark masses for $G_T=2g$}
Let us consider the thermodynamic potential at zero temperature as a function of quark
chemical potential $(\mu)$ along with the condition $F_8=F_3/\sqrt{3}$ \cite{Panda2013}. For quantitative analysis we consider the tensor coupling $G_T=2g$.
Fig.\eqref{massfig1} shows the behavior of the constituent quark masses as a function of quark chemical potential at zero temperature 
in the presence as well as in the absence of spin polarization condensate $F_3$.

In Fig.\eqref{massfig1} we have plotted the constituent quark masses as a function of quark chemical potential $\mu$ at zero temperature.
From Fig.\eqref{massfig1} it is clear that the vacuum masses ($T=0, \mu=0$), for the non strange quarks are 
0.368 GeV and the strange quark mass is 0.549 GeV. The vacuum masses for the constituent quarks remain the same as the case with $G_T=0$,
as the tensor condensates appear only at large chemical potential.
This is chiral symmetry broken phase where constituent quark masses are 
generated dynamically. Close to $\mu=\mu_c=0.360$ GeV there is  sudden drop in the masses of $u, d$ quarks $M_u=M_d$.
Because of the flavour mixing due to the determinant interaction the strange 
quark mass also changes at $\mu=\mu_c$. This sudden change in the constituent mass indicates a first-order phase transitions.
It is also expected that chiral phase transition should occur in the 
2+1 flavor NJL model near $\mu=0.360$ GeV at zero temperature in the absence of spin polarization. 
Using the gap equations it can be shown that at zero temperature and zero chemical potential $F_3=0$ is a solution. 
It turns out that 
at zero temperature and zero chemical potential $F_3=0$ is also a stable solution, hence $F_3$  does not affect the constituent 
quark masses at low chemical potential at zero temperature.
As the chemical potential is increased beyond the chiral restoration for the light quarks, it is observed that the spin polarized condensate
develop for a range of chemical potential. In particular, as shown in Fig. \eqref{fig2} for zero temperature, a non zero $F_3$ starts to develop
at $\mu\simeq 0.480$ GeV and increases slightly with $\mu$, becoming a maximum around $\mu\simeq 0.510$ GeV, beyond which it decreases
and eventually vanishes at $\mu\simeq 0.600$ GeV. Therefore we observe here that the chiral transition for the light quarks is not affected
by the spin polarization condensates as the latter exist at $\mu$ larger than $\mu_c$ for $G_T=2g$.
It is important to mention that both $\bar{\psi}\psi$
and $\bar{\psi}\gamma^{\mu}\gamma^{\nu}\psi$ break the chiral symmetry, but their thermodynamic behavior is quite opposite. 
At zero temperature and zero chemical potential non zero value of scalar condensation is thermodynamically stable, while the 
tensor condensate vanishes. However at high chemical potential when the tensor condensate takes non zero value the  
chiral condensate vanishes but for small current quark mass. As we have shown later, a strong tensor coupling can play a dominant role in
chiral phase transition and the tensor interaction can play the role of a catalyst for the chiral symmetry restoration.
The non invariance of the tensor interaction under chiral symmetry can be manifested in the change of 
quark masses even if the scalar condensate vanishes for the light quarks.

We can also understand the behavior of the
constituent quark masses $M_u=M_d$ and $M_s$ in the presence and absence of the spin polarization condensation by 
looking into the behaviour of thermodynamic potential as a function of quark-antiquark condensates $\sigma_{ud}$, $\sigma_s$ 
and spin polarization condensate
$F_3$ for different values of temperature (T) and chemical potential $\mu$. Contour plots of thermodynamic potential in the 
$\sigma_{ud}-\sigma_s$ plane for different value of chemical potential ($\mu$) at zero temperature
have been shown in Fig.\eqref{fig1} with the set of parameters given in table\eqref{table1} and $G_T=2g$.
The darker regions in the plots show the lower value of the thermodynamic
  potential. The horizontal and vertical axes represent the nonstrange quark-antiquark condensate $\sigma_{ud}$ and
  strange quark-antiquark condensate $\sigma_s$. As may be observed in Fig.\eqref{fig1}, for zero temperature and $\mu< \mu_c\sim 0.360 $ GeV
  minimization of the thermodynamic potential gives us a unique nonzero value of the quark-antiquark condensate. This 
  nonzero value of both $\sigma_{ud}$ and $\sigma_s$ indicates chiral symmetry broken phase at zero temperature 
  and $\mu\leq0.360$ GeV.
  At $\mu=0.360$ GeV one can see the existence of almost degenerate vacua in the thermodynamic potential one for $\sigma_{ud}\sim -0.015$
  GeV$^3$ and the other at $\sigma_{ud}\sim 0.0$ GeV$^3$. As the chemical potential increased this degeneracy is lifted and the
  vacuum with $\sigma_{ud}$ is close to zero has the minimum value for the thermodynamic potential. At $\mu=0.4$ GeV the value of $\sigma_{ud}$ as well as $M_u$ is very small and is close to the current quark mass value. 
  This indicates that at chemical potential larger than $\mu_c= 0.360$GeV chiral symmetry is restored. This chiral symmetry 
  restoration is partial in nature in the sence that while the scalar condensate $\sigma_{ud}\simeq 0$, but for 
  the current quark masses ($m_u,m_d\neq 0$), the strange condensate $\sigma_s$ is rather large as can be seen in 
  Fig.\eqref{massfig1} and Fig.\eqref{fig1} . As $\mu$ is further increased beyound $\mu_c$, $\sigma_s$ also approaches its (approximate)
  chiral limit continuously. Degeneracy in the 
  thermodynamic potential and a sharp jump in the order parameter ($\sigma_{ud}$) indicates first order phase transition.
  Hence the chiral transition at zero temperature is of first order in nature. 
  This first order nature of the chiral phase transition can also be seen at finite temperature, however, at relatively larger
  temperature chiral phase transition  does not remain as a first order phase transition. In fact, the end of the first order transition 
  to the crossover defines the critical end point. At higher temperatures, beyond the critical temperature quark-antiquark
  condensate changes smoothly across the critical chemical potential.
  
  When we take $G_T=2g$, the value of $F_3$ is not large enough near $\mu=0.360$ GeV and  the chiral phase transition 
 is unaffected by the spin polarization. We will qualitatively discuss the effect of $F_3$ on the chiral phase transition
by taking a relatively larger value of $G_T$ e.g. $G_T=4g$. For large value of $G_T$ the effect of spin polarization condensate 
on the chiral transition can be quite substantial. 
 Since quark-antiquark
condensates $\sigma_{ud}$ and  $\sigma_s$ are intimately connected with the $F_3$, non zero value of $F_3$ can change the quark 
dynamical mass (see Fig.\eqref{massfig1}). Strange quark mass is more affected by the presence of the spin polarization condensate $(F_3)$, because 
dynamical mass of $u$ quark becomes very small just after the chiral phase transition, however, strange quark has a substantial 
mass even after the chiral phase transition. Similar to the result at zero temperature, for $G_T=2g$ chiral phase transition 
is almost unaffected in the presence of spin polarization at finite temperature also.
For $G_T\leq 2g$  chiral phase transition is almost unaffected by the presence of the spin polarization condensate. Hence
for $G_T\leq 2g$ we have not discussed the effects of spin polarization condensate on the chiral phase transition, rather
it is important to find the domain of existence 
of  spin polarization as a function of temperature and chemical potential for different values of $G_T$ where $G_T\leq 2g$.

\subsubsection{Behavior of $F_3$ for $G_T=2g$}

Next let us focus our attention to the thermodynamic behavior of $F_3$. Fig.\eqref{fig2} shows the contour plots of the thermodynamic potential in $\sigma_s-F_3$ plane at zero temperature with 
increasing value of the chemical potential $(\mu)$ for $G_T=2g$. As before the darkest regions in the contour plots show the global minimum
of the thermodynamic potential and the corresponding values of $\sigma_s$ and $F_3$ are correct condensation value. It is clear 
from the Fig.\eqref{fig2} that spin polarization is possible within the small range of chemical potential $\mu\simeq 0.480-0.570$
GeV at zero temperature. In this work, we have kept the value of $\mu\leq\Lambda$, because $\Lambda$ is the cut-off of the theory. 
When the chemical potential is close to $0.6$ GeV both $\sigma_s$ and $F_3$ becomes zero. For large chemical potentials($\mu > $ 570 MeV),
spin polarization condensate completely melts along with the other condensates.  Presence of spin polarization condensation can affect the QCD phase diagram in many 
different ways. As we have already mentioned that the spin polarization condensate coming from the tensor interaction 
also breaks the chiral symmetry,  an obvious effect of a large value of spin polarization condensate should be seen in the 
chiral phase transition. We have also observed that $F_3$ decreases with increasing temperature and vanishes at few tens of MeV. Therefore such condensates do not affect the critical end point. 

\begin{figure}[!h] 
    \begin{minipage}[b]{0.5\linewidth}
    \centering
    \includegraphics[width=0.95\linewidth]{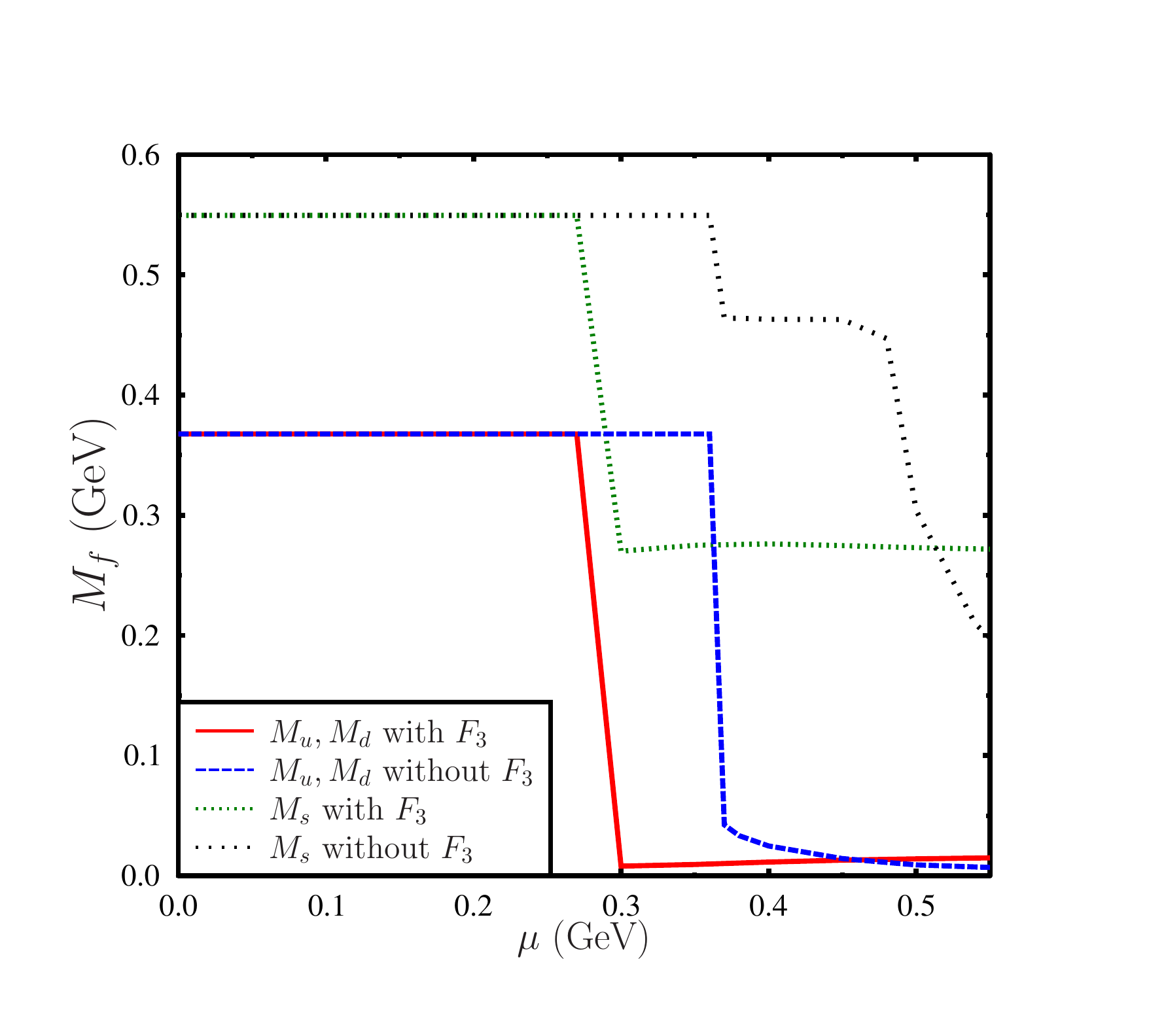} 
    \vspace{4ex}
  \end{minipage}
  \begin{minipage}[b]{0.5\linewidth}
    \centering
    \includegraphics[width=0.95\linewidth]{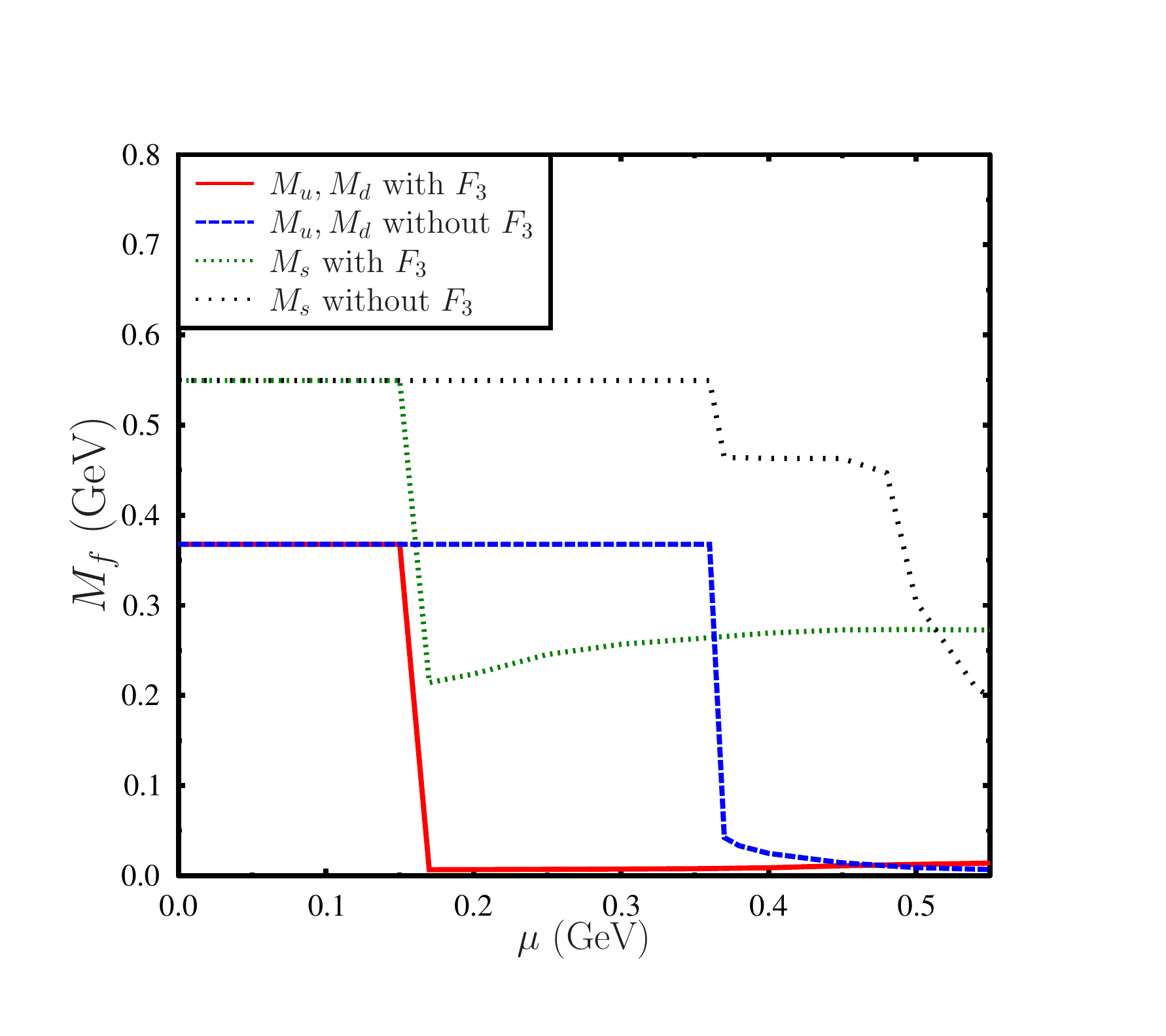} 
    \vspace{4ex}
  \end{minipage} 
\caption{Dependence of constituent quark mass on the quark chemical potential at zero temperature in the presence
 as well as in the absence of 
 spin polarization condensation for different values of tensor couplings $G_T=4g$ (left plot) and $G_T=4.3g$ (right plot)
 for $F_8=F_3/\sqrt{3}$.
 Red-solid line and green-dotted line represent non strange and strange quark mass in the
 presence of spin polarization condensate $F_3$. Blue-dashed line  and black-dotted line represents non strange and strange quark
 constituent mass in the standard 2+1 flavor NJL model in the absence of any spin polarization condensate. Sharp jump in the 
 value of $M_u$ and $M_s$ near $\mu=0.360$ GeV in both plots indicates the first order chiral phase transition which is 
 expected for standard 2+1 flavour NJL model. From this plots it is clear that for larger value of the tensor coupling 
 the chiral phase transition occurs at a smaller value of chemical potential. For larger tensor coupling tensor condensate form 
 at relatively smaller chemical potential and it remains non zero for a wide range of chemical potential. This large value of the spin 
 polarization condensate even for large chemical potential affects the quark masses substantially.}
  \label{massfig2}
\end{figure}
\subsubsection{Chiral phase transition and the behavior of quark masses for larger tensor coupling}
The left plot and the right plot in Fig.\eqref{massfig2} are for the tensor coupling $G_T=4g$ and $G_T=4.3g$ respectively. In 
Fig.\eqref{massfig2} one can see the effects of the large tensor couplings on the chiral phase transition as well as on the 
quark masses. For $G_T=4g$ and $G_T=4.3g$ the chiral phase transition occurs
at $\mu=0.270$ GeV and $\mu=0.170$ GeV respectively, which are lower than the critical chemical potential in the absence of any 
spin polarization at zero temperature. It is also important to see the effect of large tensor coupling on the quark masses. For 
large tensor coupling spin polarization condensate has non zero value for a wide range of chemical potential. Since the chiral 
condensates are intimately connected with the spin polarization condensate any non zero value of the spin polarization also 
affects the quark masses in the chiral symmetry restored phase. It is important to note that non zero value of $F_3$
has larger effect on the strange quark mass rather than the non strange quark masses, as may be observed in Fig. \eqref{massfig1}.

\begin{figure}[!h] 
    \minipage{0.33\linewidth}
    \includegraphics[width=\linewidth]{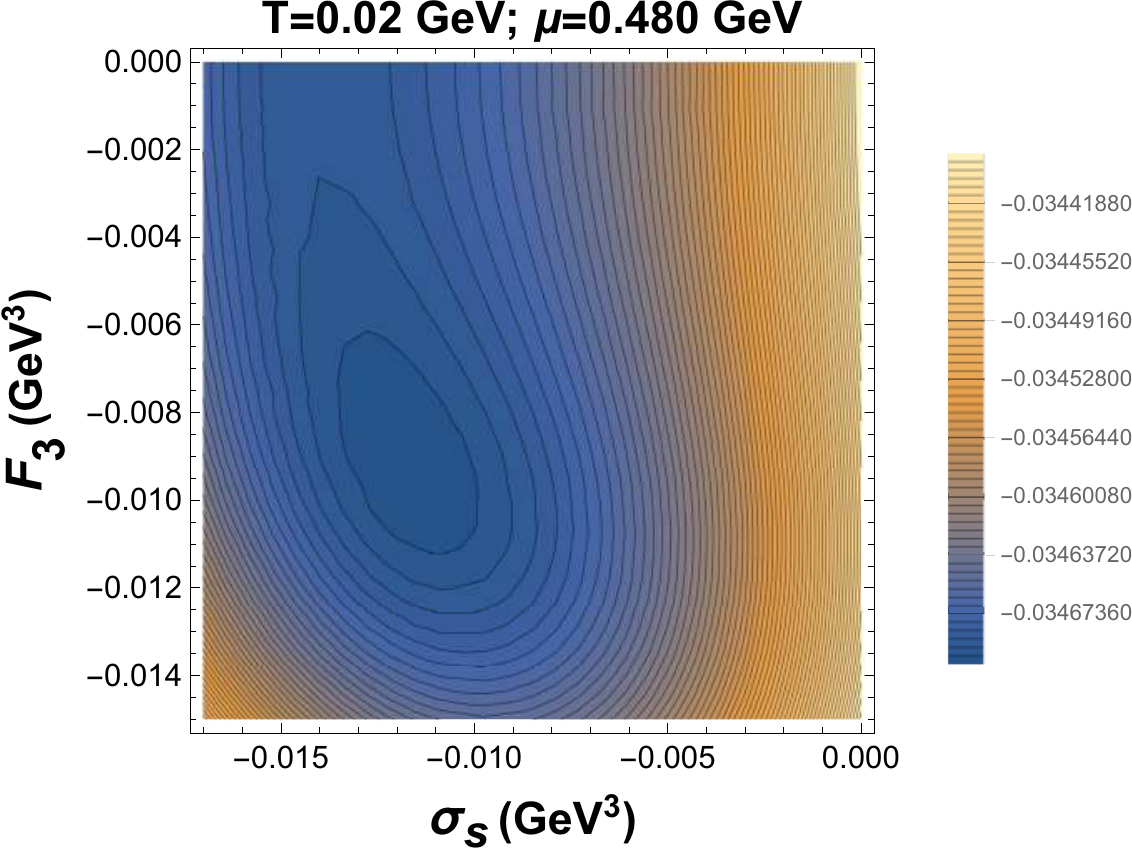} 
     \vspace{4ex}
  \endminipage\hfill
  \minipage{0.33\linewidth}
    \includegraphics[width=\linewidth]{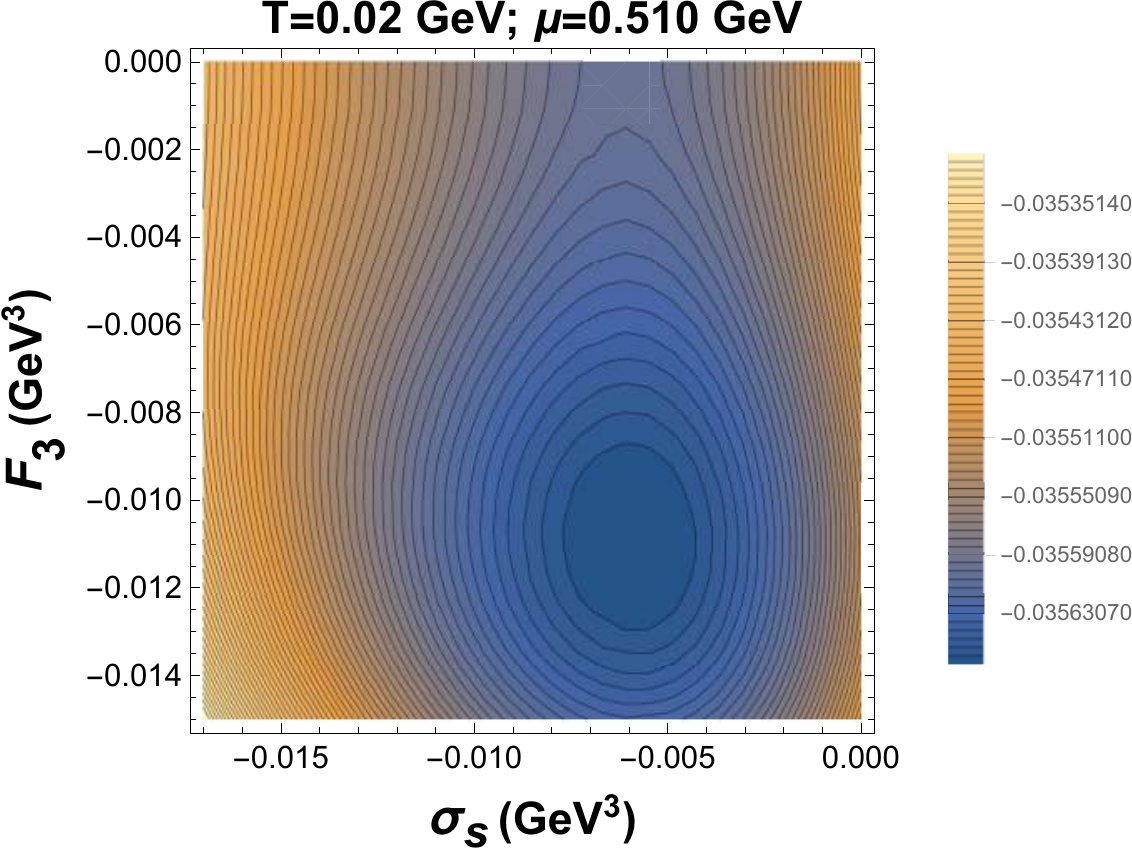} 
     \vspace{4ex}
  \endminipage\hfill
  \minipage{0.33\linewidth}
    \includegraphics[width=\linewidth]{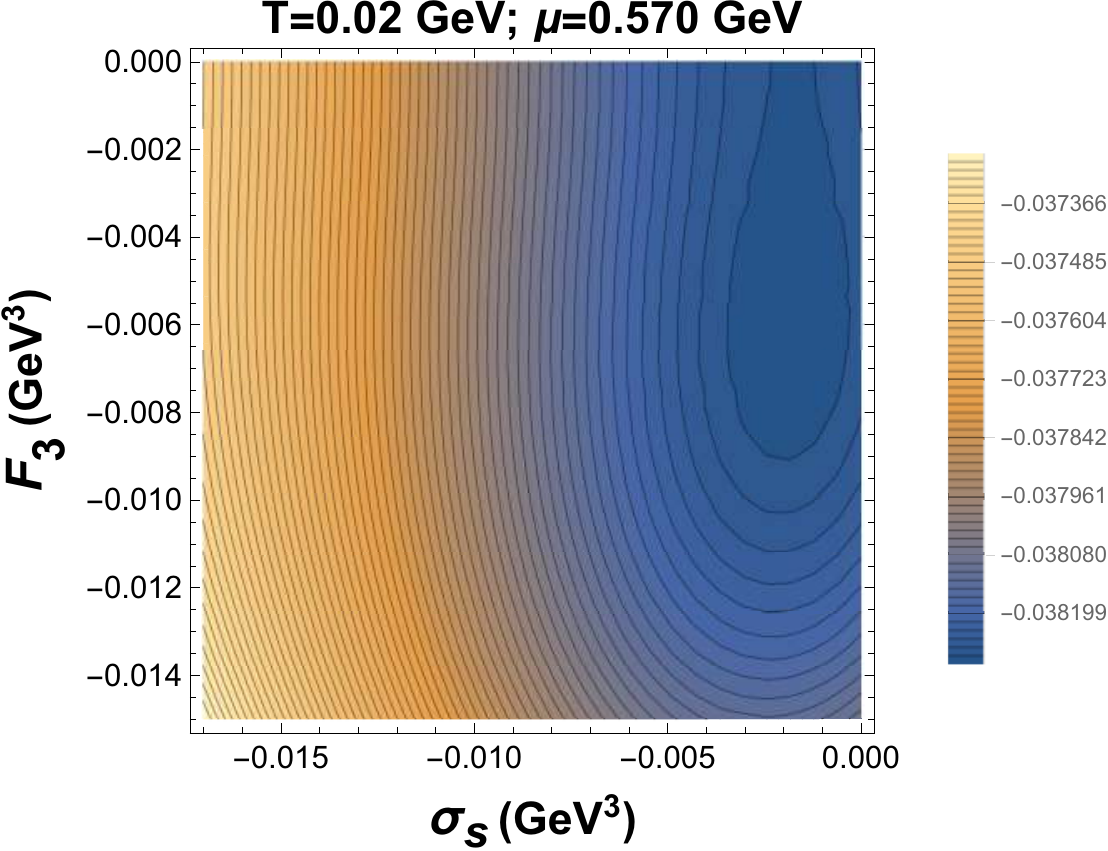} 
     \vspace{4ex}
  \endminipage\hfill
  \minipage{0.33\linewidth}
    \centering
    \includegraphics[width=\linewidth]{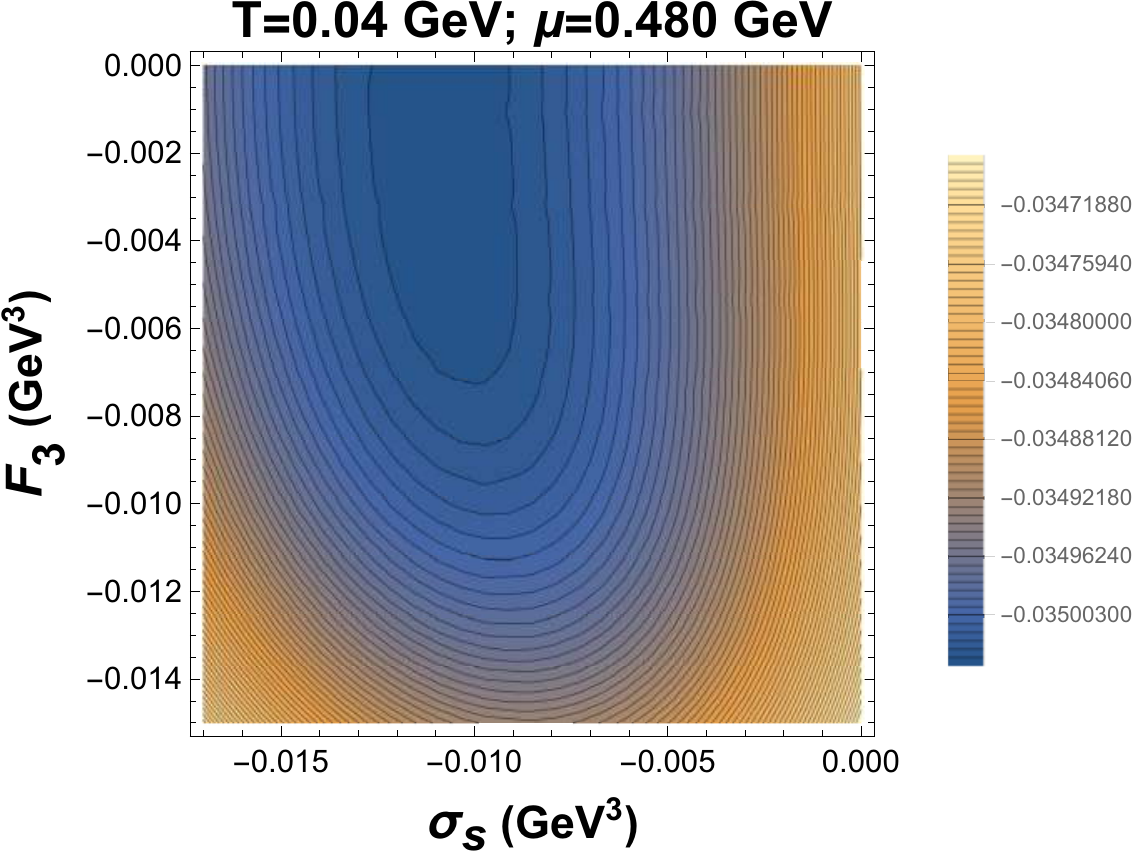} 
    \vspace{4ex}
  \endminipage\hfill
  \minipage{0.33\linewidth}
    \includegraphics[width=\linewidth]{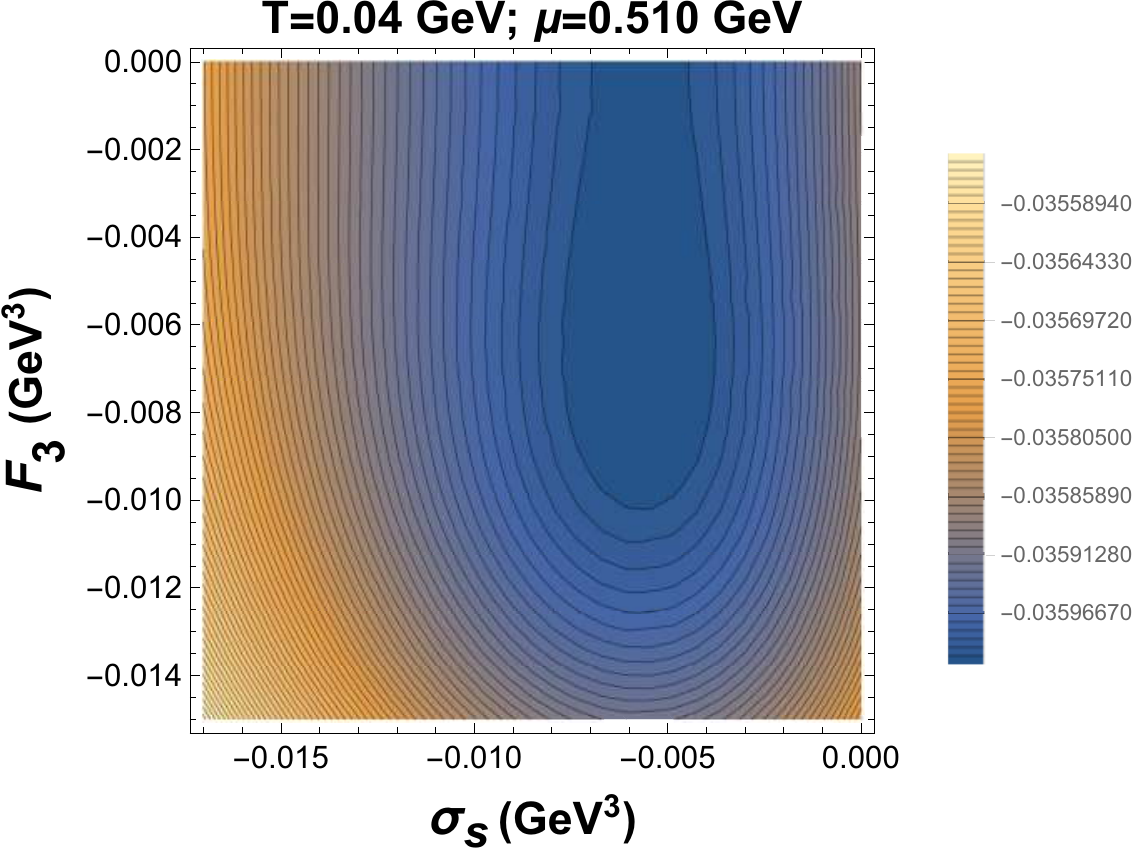} 
    \vspace{4ex}
  \endminipage\hfill 
  \minipage{0.33\linewidth}
    \includegraphics[width=\linewidth]{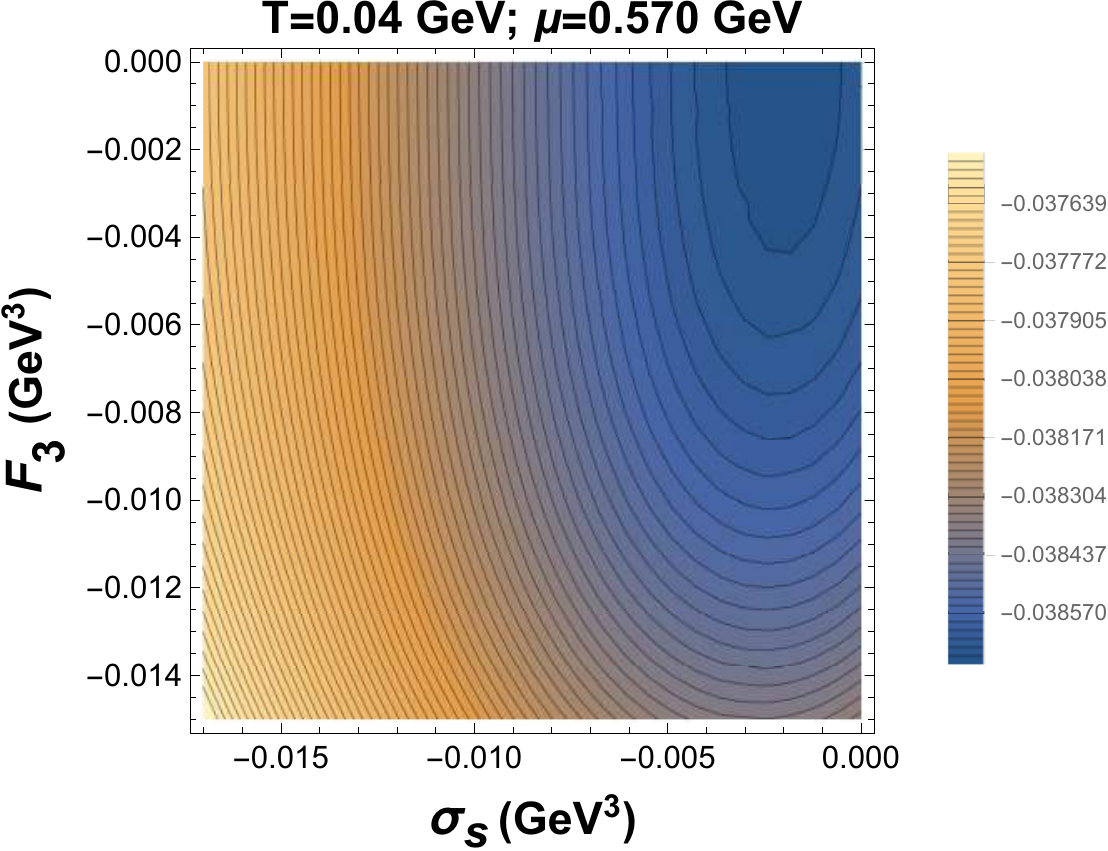} 
     \vspace{4ex}
  \endminipage\hfill 
   \minipage{0.33\linewidth}
    \includegraphics[width=\linewidth]{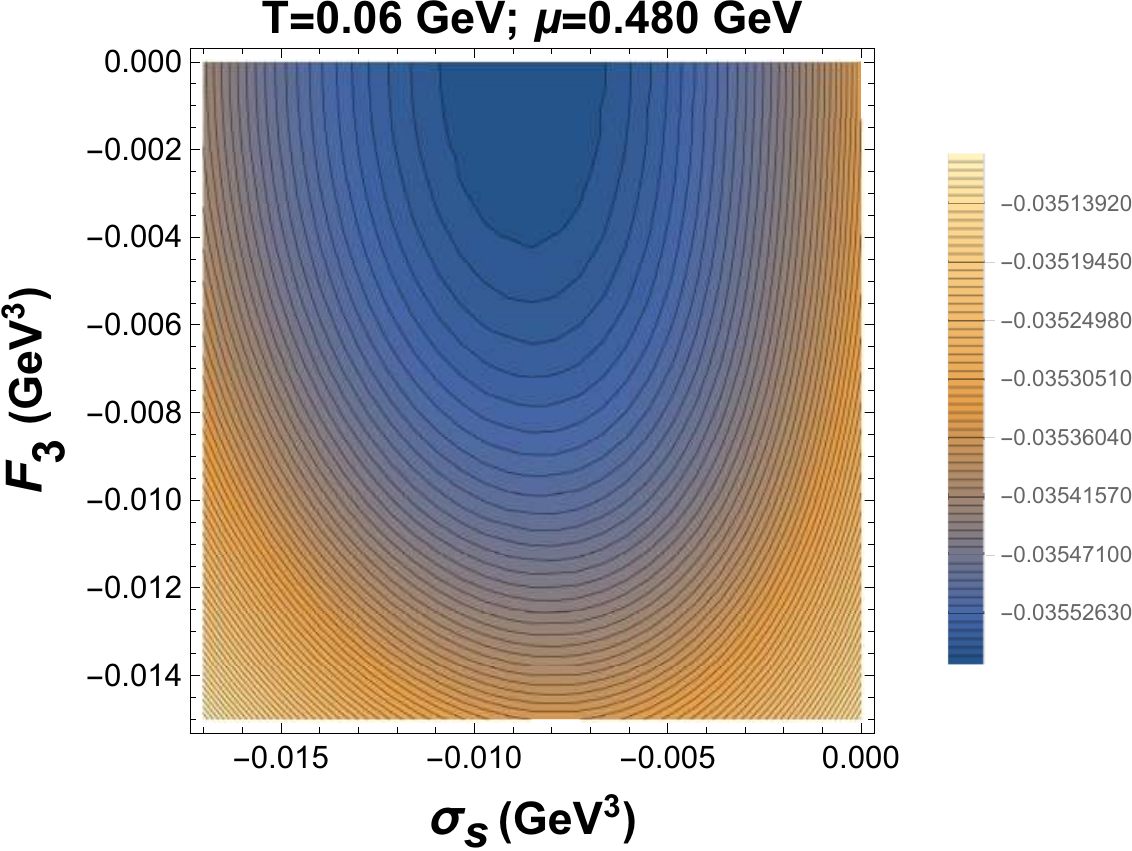} 
     \vspace{4ex}
  \endminipage\hfill 
   \minipage{0.33\linewidth}
    \includegraphics[width=\linewidth]{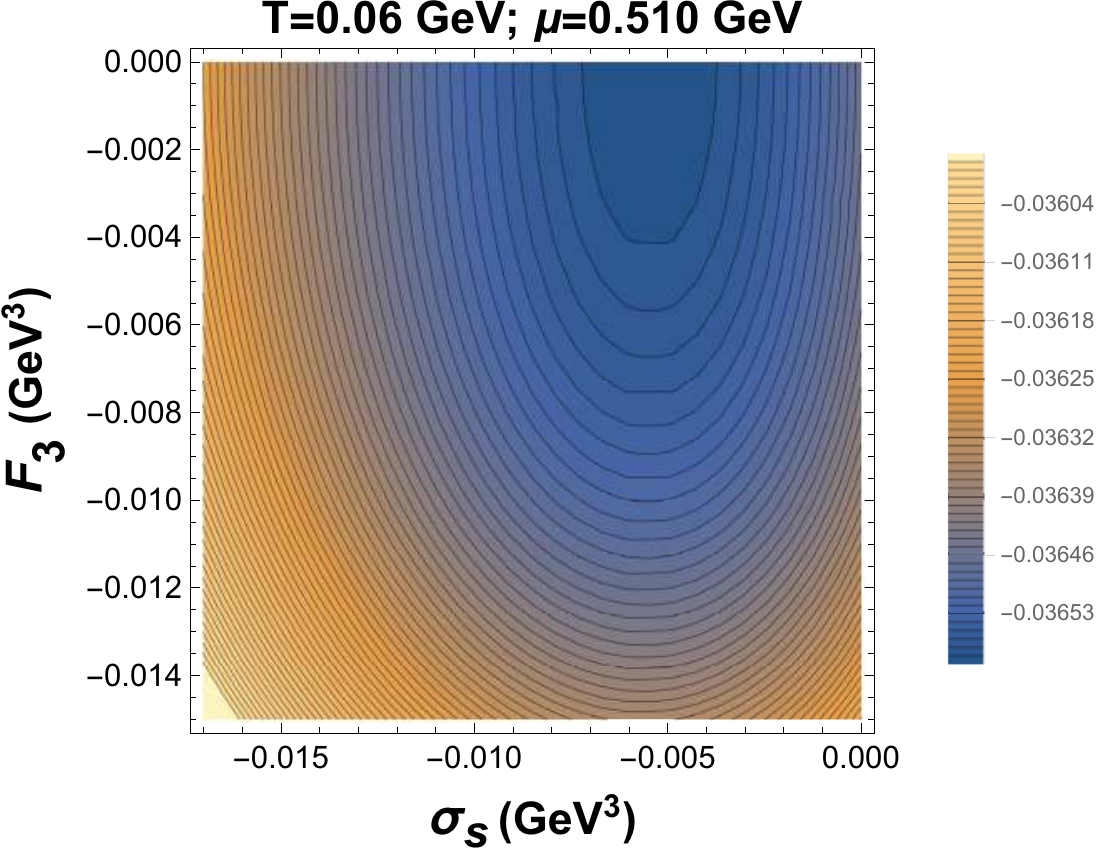} 
     \vspace{4ex}
  \endminipage\hfill 
   \minipage{0.33\linewidth}
    \includegraphics[width=\linewidth]{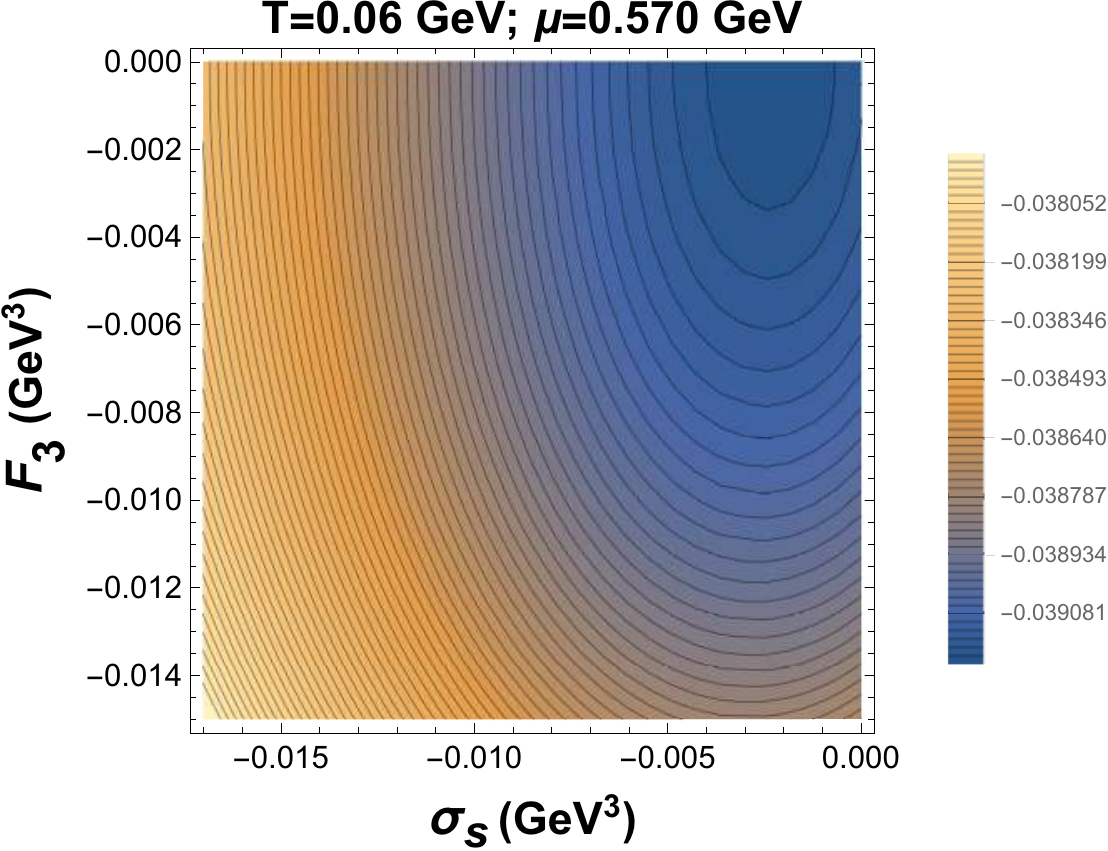} 
     \vspace{4ex}
  \endminipage\hfill 
  \caption{This figure shows the contour plots of the thermodynamic potential in $\sigma_s-F_3$ plane for finite temperature (T)
  and finite chemical potential ($\mu$) with $G_T=2g$ and $F_8=F_3/\sqrt{3}$. Along each row as we move from left to the right,
  temperature has been kept fixed but
  $\mu$ is increasing, similarly along each column $\mu$ has been kept fixed with $T$ increasing. Darker regions in these contour 
  plots show the global minimum of the thermodynamic potential. It is clear from the plots that at small temperature non zero 
  value of the spin polarization starts to appear at smaller value of the chemical potential and it also melts at higher chemical
  potential. Thus for smaller temperature the domain of $\mu$ where one can get non zero spin polarization is larger. This domain 
  of existence for the spin polarization condensate becomes smaller with increasing temperature $T$ for a given value of $G_T$. In fact when
  the temperature is $T=0.06$ GeV we cannot get spin polarization for any value of $\mu$.}
  \label{fig3}
\end{figure}
\subsubsection{Finite temperature effect on the spin polarization condensate $F_3$ for $G_T=2g$}
After demonstrating the behavior of the spin polarization condensate as a function of chemical potential at zero temperature
for different values of the tensor coupling, let us look into
the temperature
behavior of $F_3$ for a fixed value of $G_T=2g$. Temperature behavior of spin polarization condensate as well as $\sigma_s$ is shown in Fig.\eqref{fig3}.
 Fig.\eqref{fig3} shows the contour plots of thermodynamic potential in the plane of $\sigma_s-F_3$ for different values of
 temperature and chemical potential. Each row shows the behavior of thermodynamic potential as a function of increasing
 chemical potential for a fixed temperature. On the other hand, each column shows the behavior of the thermodynamic potential 
 as a function of temperature for a fixed value of chemical potential. From the first two row in Fig.\eqref{fig3}, for temperature
 $T=0.02$ GeV and $0.04$ GeV, it is clear that as the chemical potential increase non zero value of spin polarization develops.
 It attains some maximum value at an intermediate value of the chemical potential and as the chemical potential becomes very high 
 $F_3$ becomes zero. However, each column shows that with increasing temperature the formation of the spin
 polarization becomes difficult and the maximum value of $F_3$ also decreases with temperature. The third row 
 in Fig.\eqref{fig3} shows that 
 when the temperature is $T=0.06$ GeV, value of the spin polarization condensate $F_3$ is almost zero. Hence
 one can conclude that as the temperature increases the range of chemical potential within which spin polarization 
 can exist decreases. Further there exists a temperature beyond which spin polarization cannot occur irrespective of
 the value of chemical potential for a given value of $G_T$.

\subsubsection{Threshold coupling for existence of $F_3$} 
\begin{figure}[!h] 
    \minipage{0.33\linewidth}
    \includegraphics[width=\linewidth]{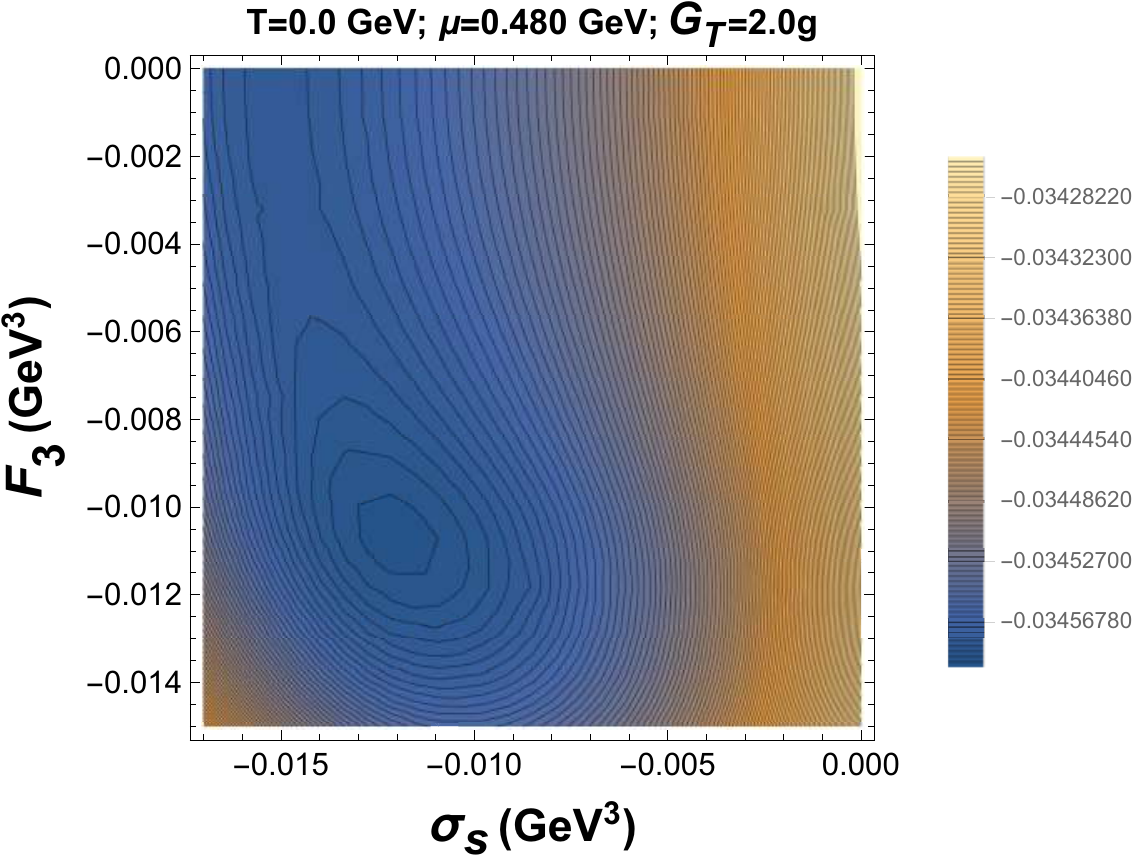} 
     \vspace{4ex}
  \endminipage\hfill
  \minipage{0.33\linewidth}
    \includegraphics[width=\linewidth]{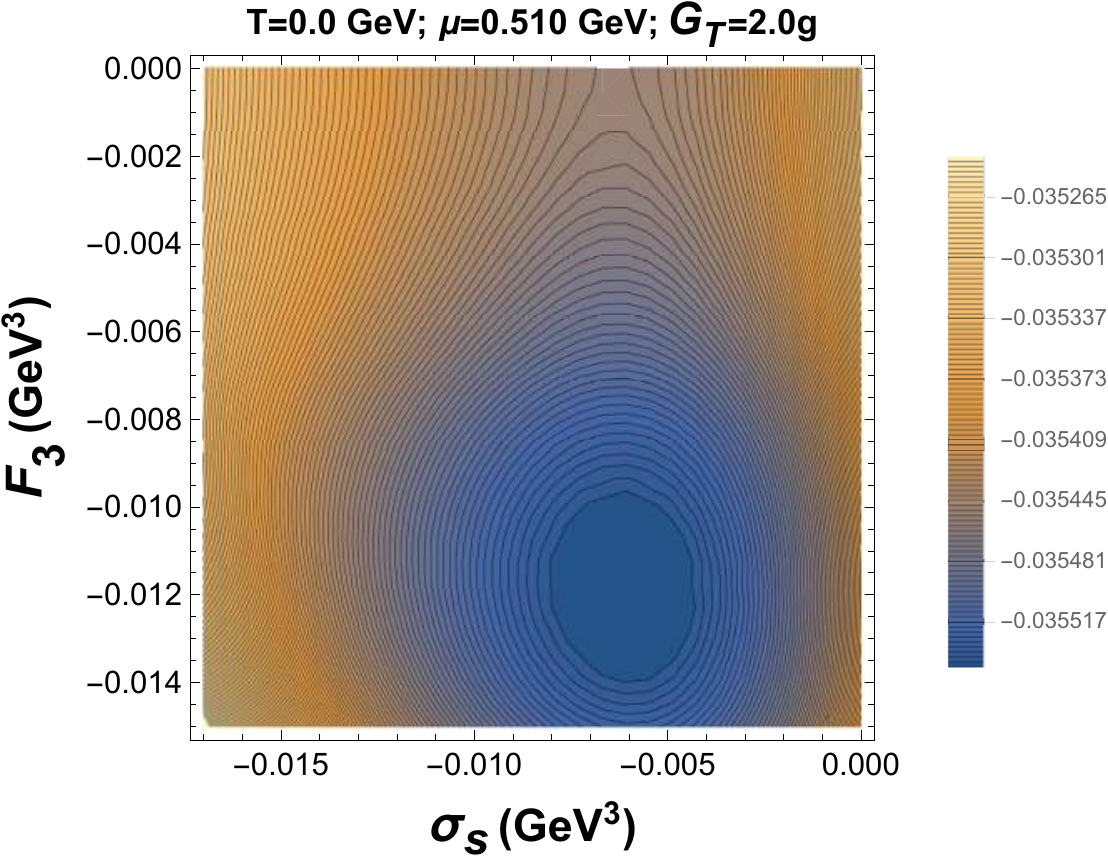} 
     \vspace{4ex}
  \endminipage\hfill
  \minipage{0.33\linewidth}
    \includegraphics[width=\linewidth]{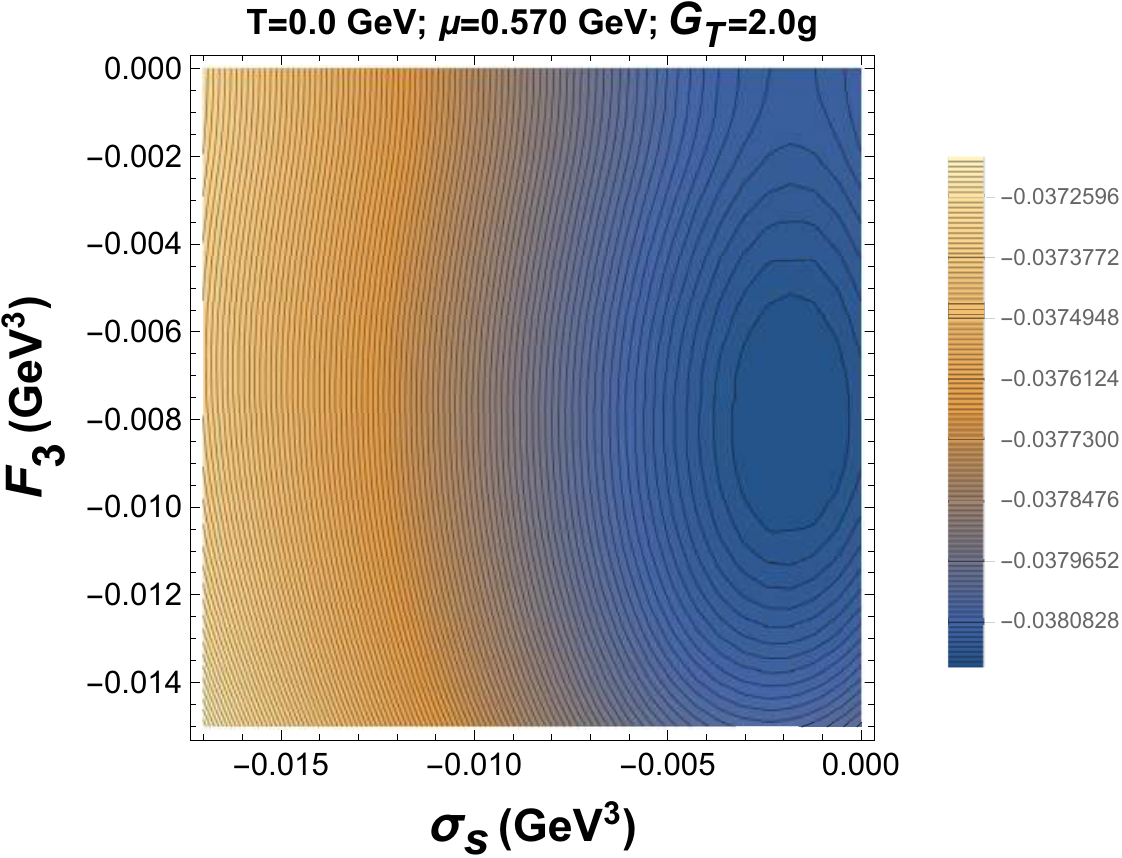} 
     \vspace{4ex}
  \endminipage\hfill
  \minipage{0.33\linewidth}
    \centering
    \includegraphics[width=\linewidth]{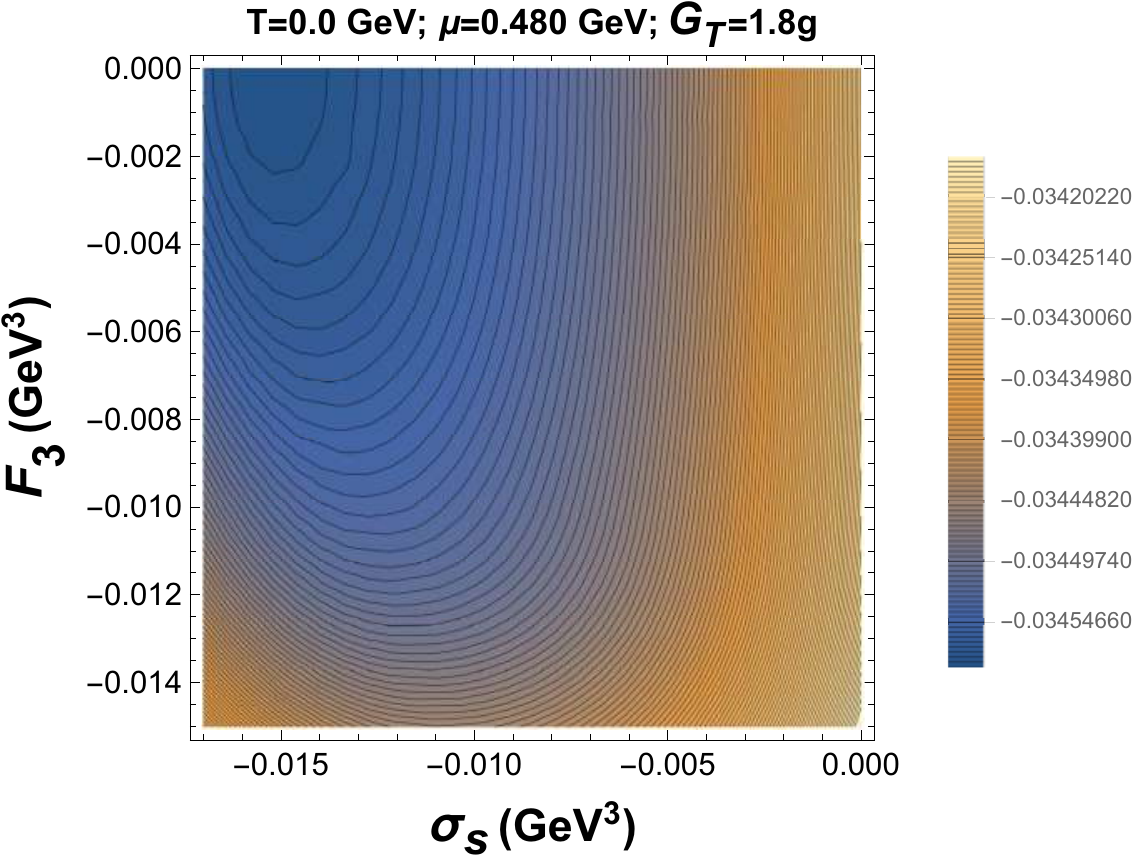} 
    \vspace{4ex}
  \endminipage\hfill
  \minipage{0.33\linewidth}
    \includegraphics[width=\linewidth]{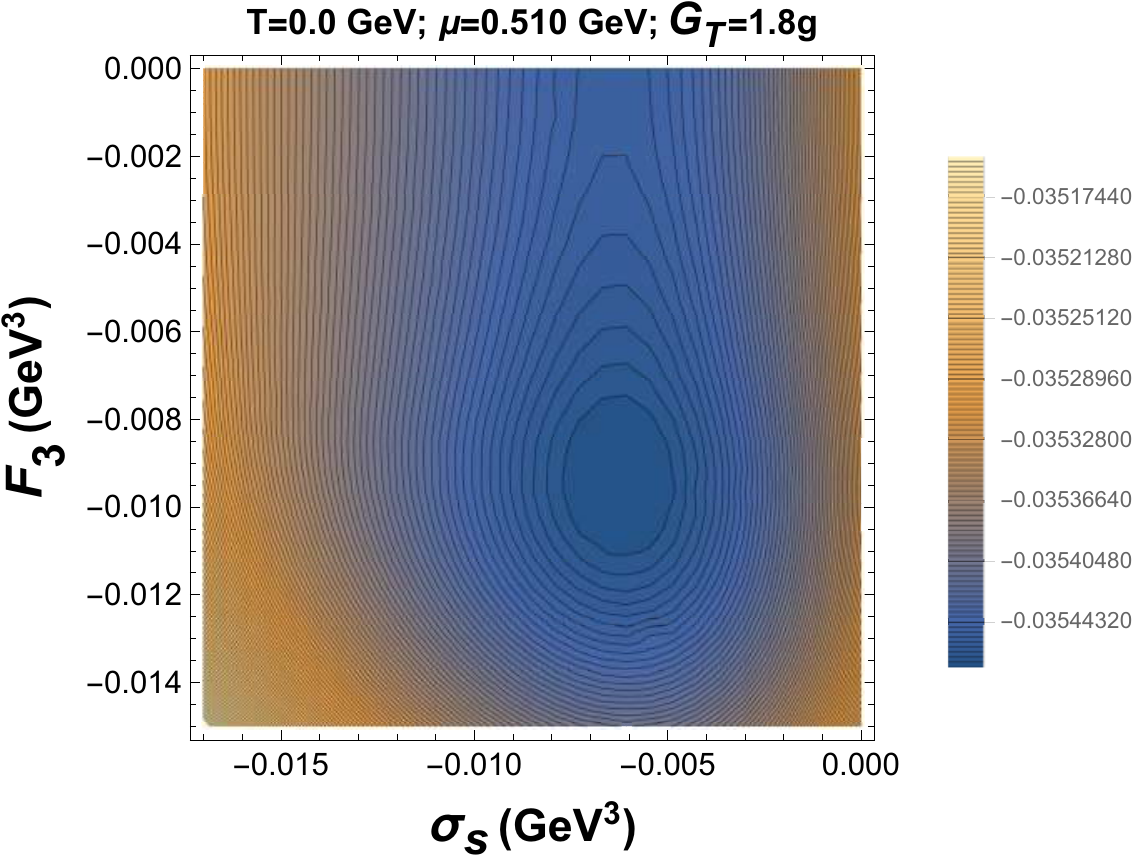} 
    \vspace{4ex}
  \endminipage\hfill 
  \minipage{0.33\linewidth}
    \includegraphics[width=\linewidth]{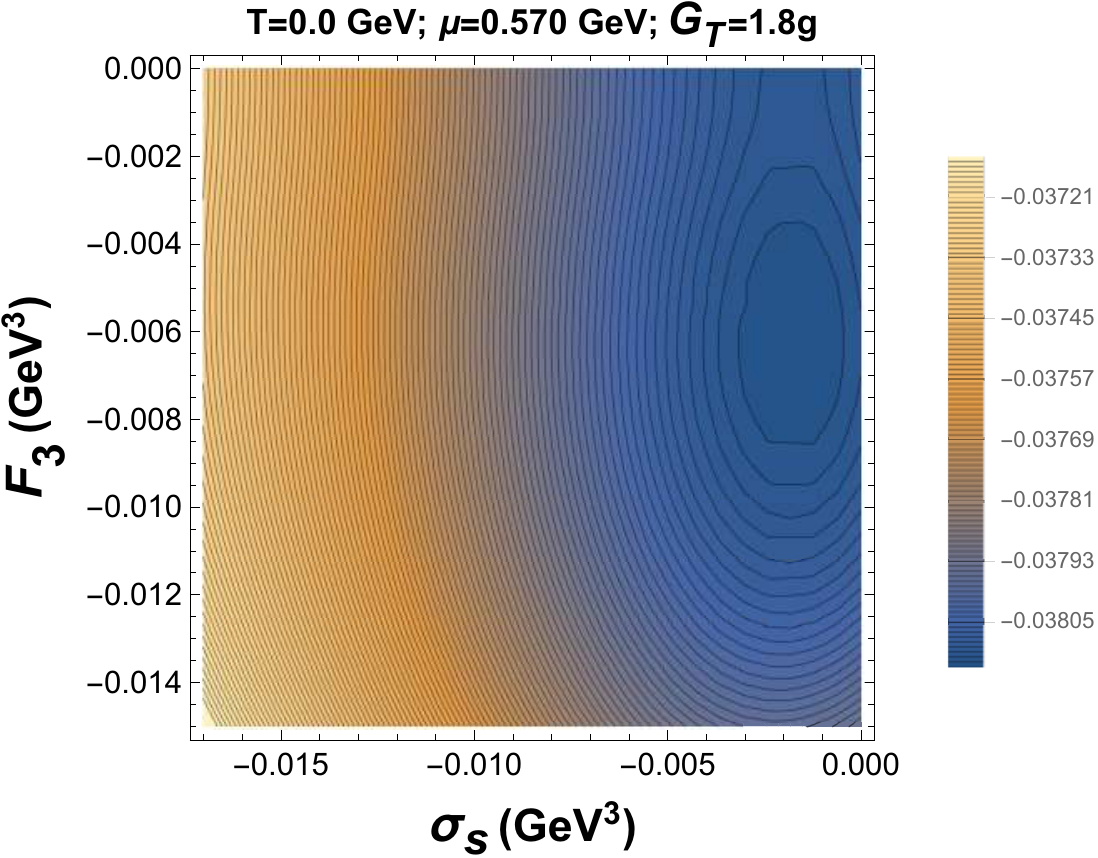} 
     \vspace{4ex}
  \endminipage\hfill 
   \minipage{0.33\linewidth}
    \includegraphics[width=\linewidth]{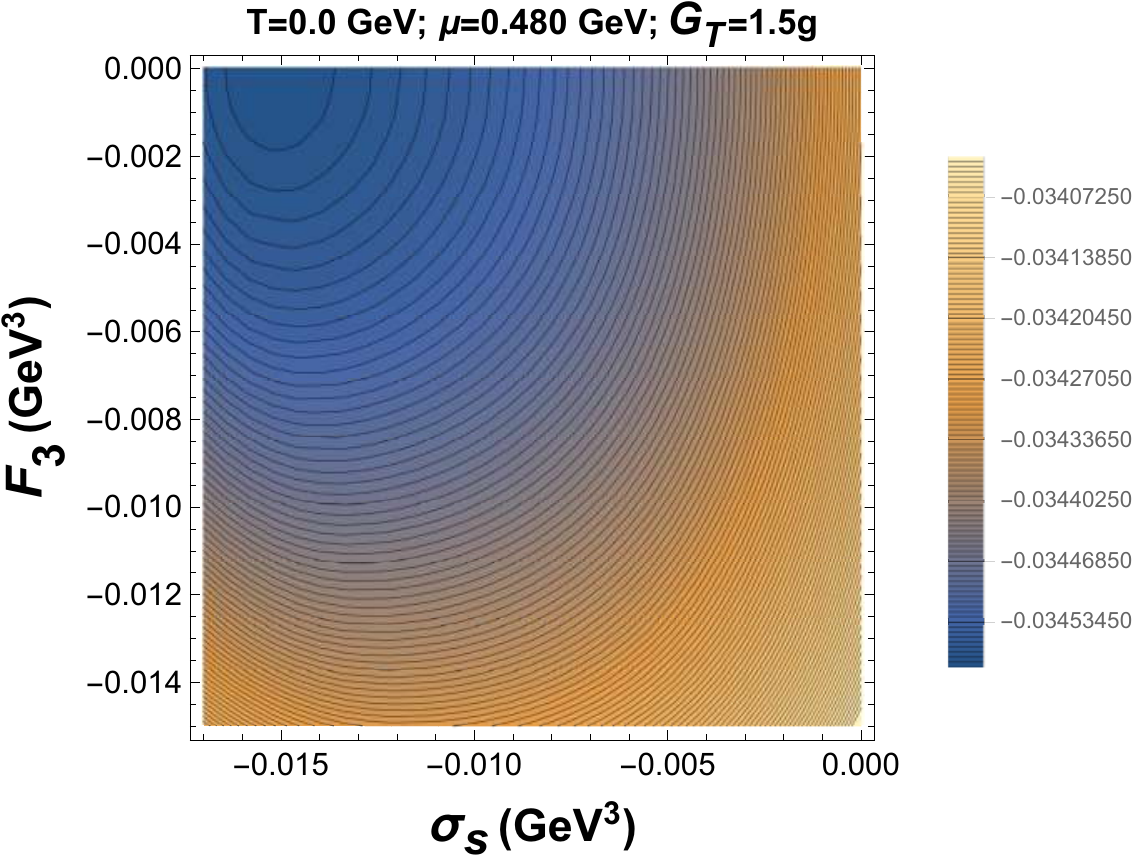} 
     \vspace{4ex}
  \endminipage\hfill 
   \minipage{0.33\linewidth}
    \includegraphics[width=\linewidth]{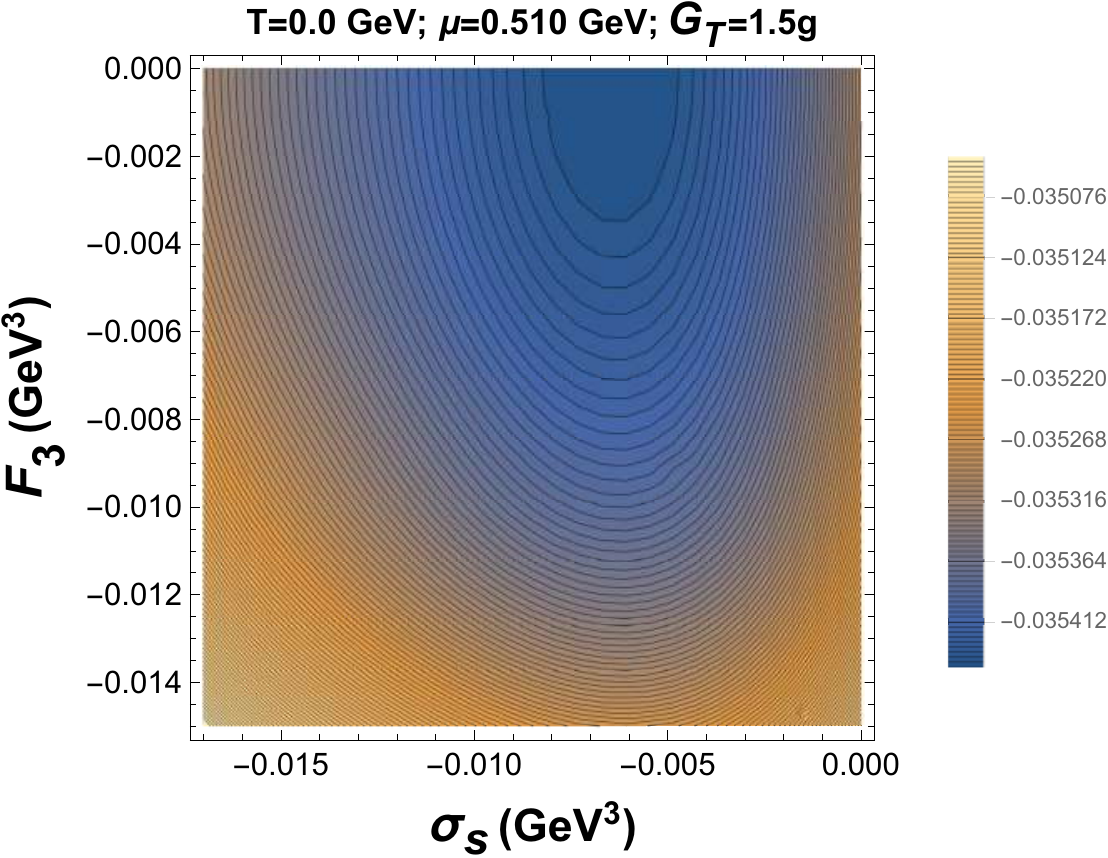} 
     \vspace{4ex}
  \endminipage\hfill 
   \minipage{0.33\linewidth}
    \includegraphics[width=\linewidth]{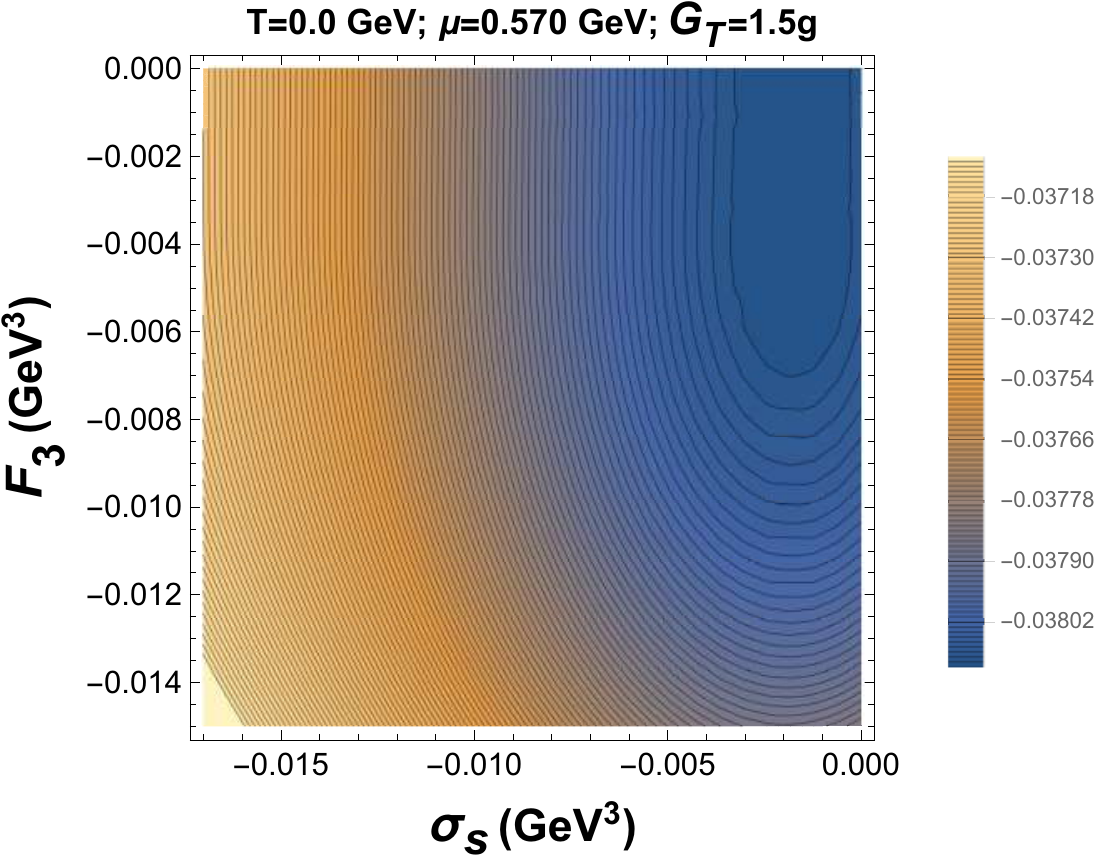} 
     \vspace{4ex}
  \endminipage\hfill 
  \caption{This figure shows the contour plots of the thermodynamic potential in $\sigma_s-F_3$ plane for zero temperature (T)
  and finite chemical potential ($\mu$) with different values of tensor coupling $G_T$ and $F_8=F_3/\sqrt{3}$. In the 
  first, second and the third row 
  the tensor couplings are taken as $G_T=2g,1.8g$ and $1.5g$ respectively. Along each row temperature and $G_T$ has been 
  kept fixed but $\mu$ is increasing, similarly along each column $\mu$ and $T$ has been kept fixed with $G_T$ decreasing.
  Darker regions in these contour plots shows the global minimum of the thermodynamic potential. 
  It is clear from the plots that at zero temperature, for larger value of tensor coupling  spin polarization can exist for a
  relatively wide range of chemical potential. With the decreasing value  of tensor coupling  e.g. for $G_T=1.5g$ spin polarization almost
  vanishes. This result can be easily extended to finite temperature. For non zero temperature existence of spin polarization
  requires lager value of $G_T$.}
  \label{fig4}
\end{figure}

The existence of spin polarization inevitably depends on the value of $G_T$. $G_T$ determines the strength of
the spin polarization condensation. The dependence of $F_3$ on the tensor coupling has been shown in
the Fig\eqref{fig4}. Fig.\eqref{fig4} shows 
the thermodynamic potential in $\sigma_s-F_3$ plane as  a function of chemical potential for three different
values of tensor 
couplings $G_T=2g,1.8g$ and $1.5g$ at zero temperature. Along each row in Fig.\eqref{fig4} the contours of thermodynamic
potential
have been shown for different values of the chemical potential but keeping $G_T$ fixed. On the other hand in each 
column of Fig.\eqref{fig4}
contours of thermodynamic potential have been shown for various values the tensor coupling constant $G_T$ for a given chemical potential.
Value of the spin polarization condensate decreases with decreasing value of $G_T$.
When $G_T=2g$, $F_3$ has a substantial non zero value at zero temperature and $\mu=0.510$ GeV,
however for $G_T=1.8g$ this value starts to decrease and for $G_T=1.5g$  spin polarization condensate $F_3$ almost vanishes.
This result for zero temperature
can be easily extended to a non zero temperature. For finite temperature one requires a larger value of $G_T$, for the
spin polarization to exist. As $G_T$ increases, the threshold $\mu$ above which $F_3$ starts becoming nonvanishing decreases, and the critical $\mu$ above which $F_3$ vanishes increases.
Both these behavior lead to a larger range of $\mu$ that supports a non vanishing $F_3$ as $G_T$ increases. Further the magnitude of $F_3$ increases with $G_T$.  
 
 \subsection{Results for independent $F_3$, $F_8$}
 \subsubsection{Thermodynamics behavior of $F_3$ and $F_8$ separately for $G_T=2g$}
\begin{figure}[!h] 
    \begin{minipage}[b]{0.5\linewidth}
    \centering
    \includegraphics[width=0.95\linewidth]{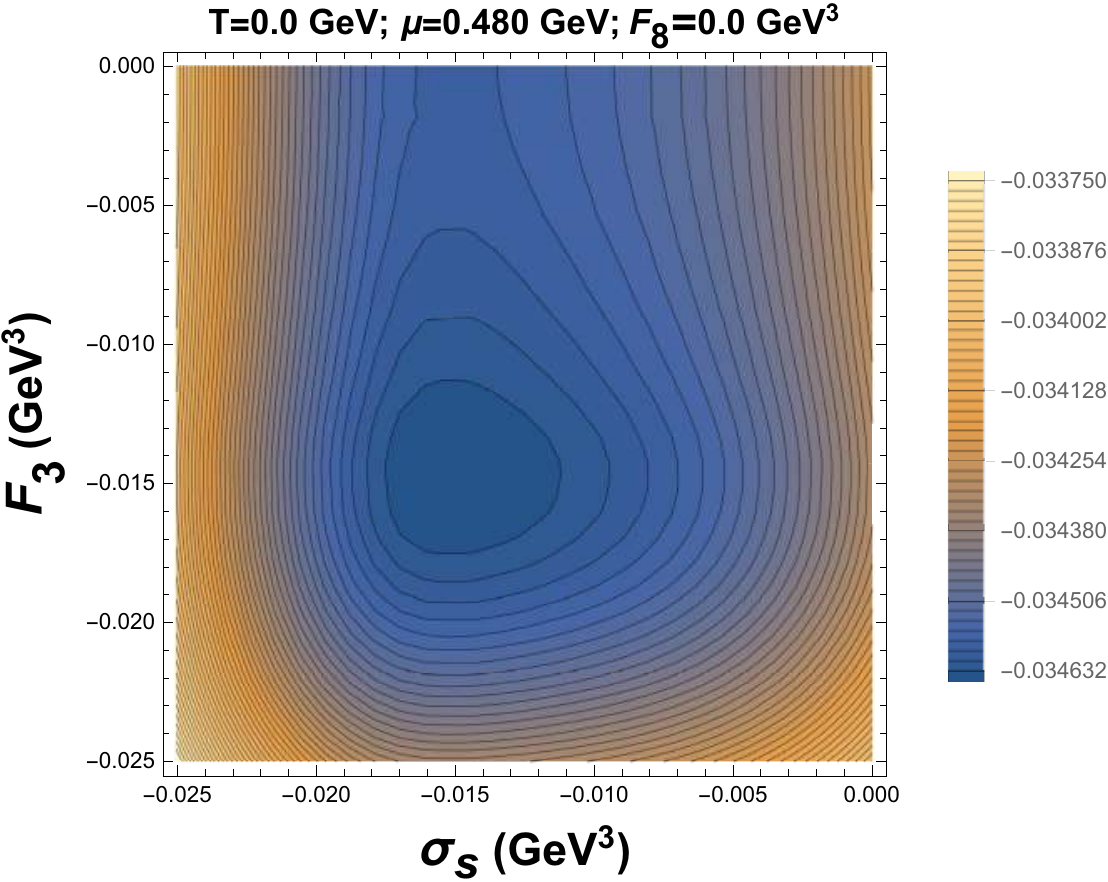} 
    \vspace{4ex}
  \end{minipage}
  \begin{minipage}[b]{0.5\linewidth}
    \centering
    \includegraphics[width=0.95\linewidth]{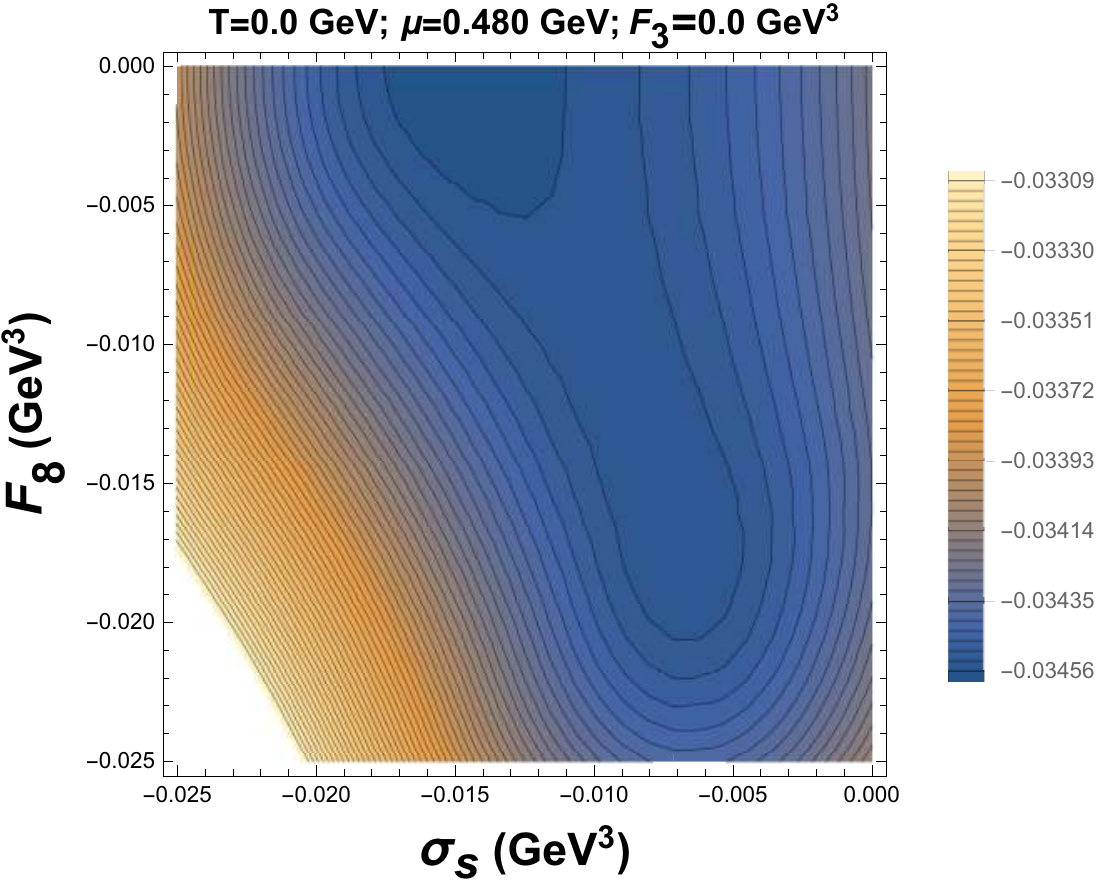} 
    \vspace{4ex}
  \end{minipage} 
  \begin{minipage}[b]{0.5\linewidth}
    \centering
    \includegraphics[width=0.95\linewidth]{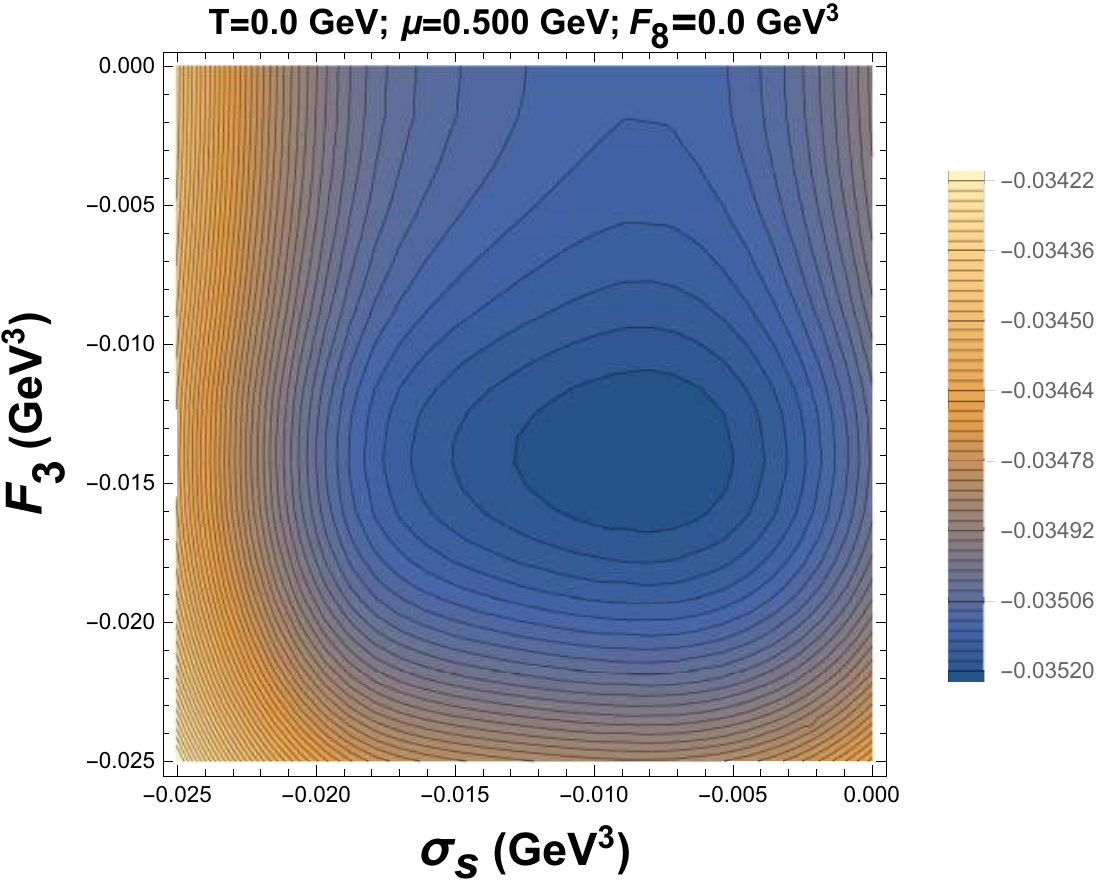} 
    \vspace{4ex}
  \end{minipage}
  \begin{minipage}[b]{0.5\linewidth}
    \centering
    \includegraphics[width=0.95\linewidth]{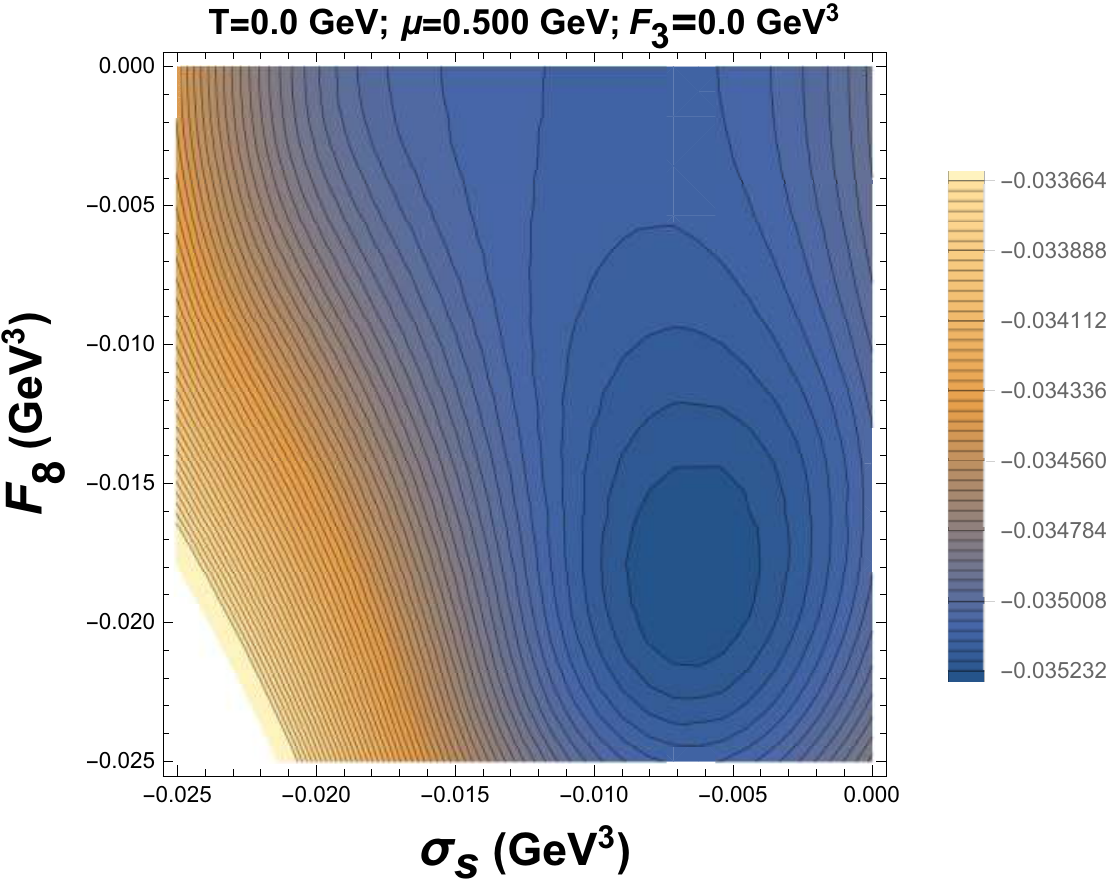} 
    \vspace{4ex}
  \end{minipage}
  \begin{minipage}[b]{0.5\linewidth}
    \centering
    \includegraphics[width=0.95\linewidth]{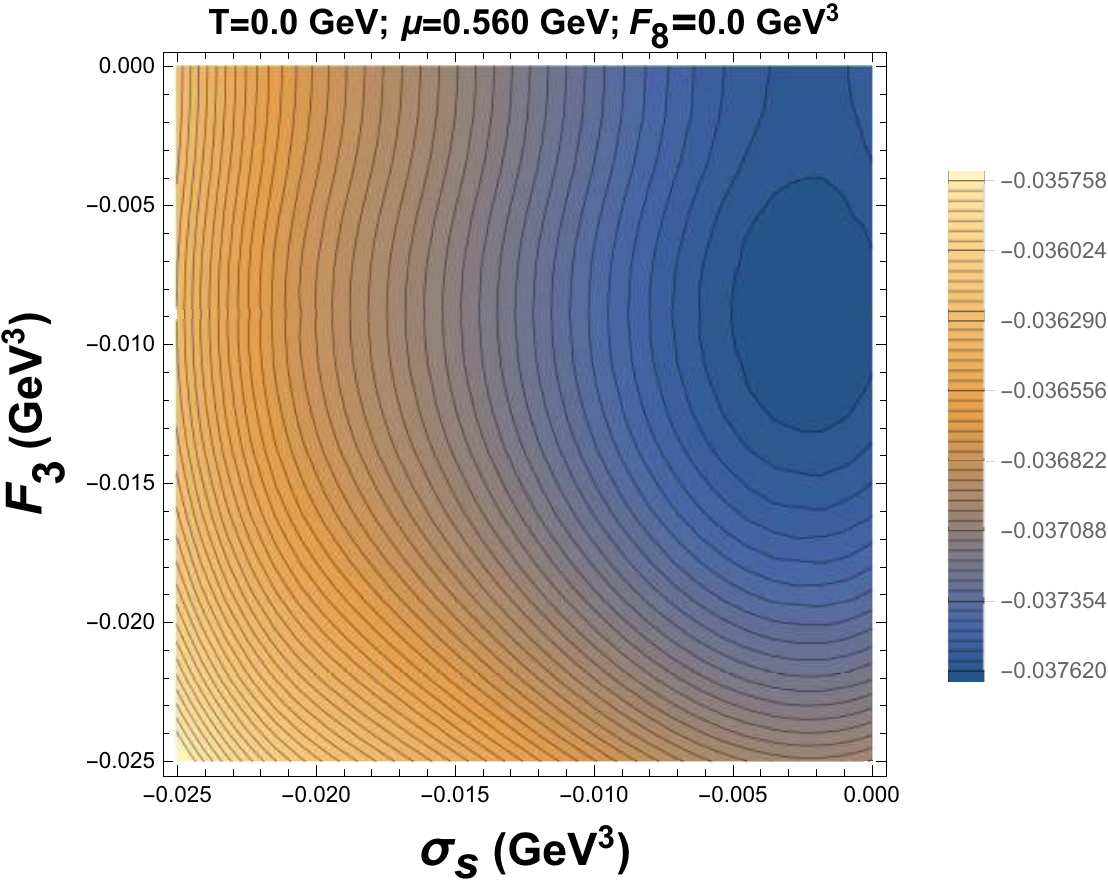} 
    \vspace{4ex}
  \end{minipage}
  \begin{minipage}[b]{0.5\linewidth}
    \centering
    \includegraphics[width=0.95\linewidth]{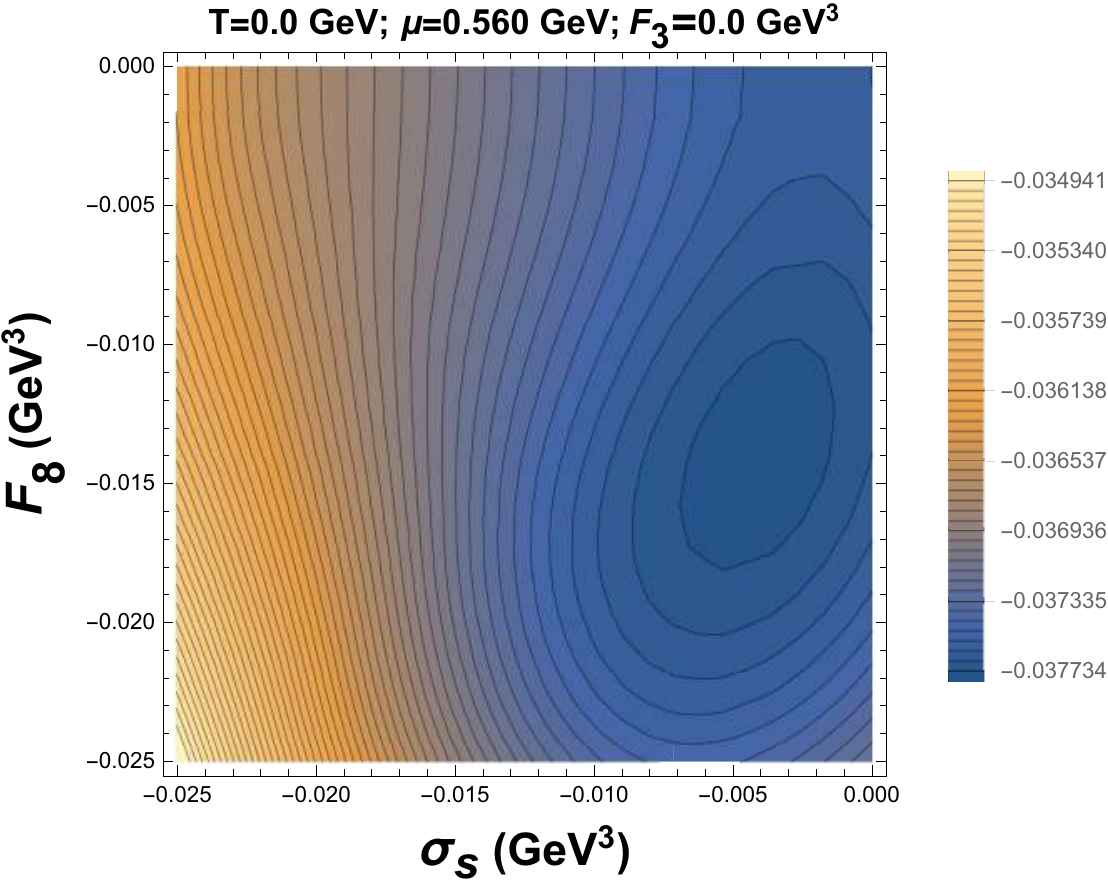} 
    \vspace{4ex}
  \end{minipage} 
  \caption{This figure shows the behaviour of thermodynamic potential where $F_3$ and $F_8$ are considered independently, for $G_T=2g$.
  In the first column only 
  the effect of $F_3$ is considered and the in the second column only $F_8$ is considered. In this figure contour plots of the thermodynamic
  potential in the $\sigma_s-F_3$ plane and in the $\sigma_s-F_8$ plane at zero temperature and finite chemical potential has been shown.
  Along each row the quark 
  chemical potential has been kept constant. From this plot it is clear that spin polarization of type $F_3$ occurs for a relatively small $\mu$
  with respect to $F_8$ and $F_3$ also melts earlier than $F_8$ as we increase the chemical potential.}
  \label{fig7}
\end{figure}

 In the earlier discussions we have considered a simplified approximation where $F_8=F_3/\sqrt{3}$, leading to 
 $\langle\bar{d}\Sigma_z d\rangle = \langle\bar{s}\Sigma_z s\rangle $ for the sake of simplicity. However the masses of the 
 light and the strange quarks are different, the $d$ and $s$ quark spin polarization condensate need not be at the same footing.
 This is the scenario that we wish to explore here for completeness. 
 In Fig.\eqref{fig7}, again to investigate the behavior of light and the strange quark spin polarized condensates we have first
 considered $F_8$=0 and studied the behavior  of the thermodynamic potential as a function of $F_3$ only. This is 
 explained as shown in the left panel of Fig.\eqref{fig7}. The spin polarization condensate for the light quarks begin to
 develop for $\mu$= 0.48 GeV. On the right panel of Fig.\eqref{fig7}we have plotted the figure but taken $F_3$=0 and
 considered the thermodynamic potential as a function of $F_8$ only. Let us recall that while $F_3$ involves
 the difference between spin polarization condensates of two light quarks, the $F_8$ involves the difference
 of spin polarization condensates of light quarks and the strange quark. Thus the left panel corresponds to the
 case when sum of the light quark spin condensate is equal to the strange quark spin polarization condensate and
 the right panel on the other hand corresponds to the case when the two light quark spin polarization condensates
 are equal. It is observed that for the latter case ($F_8 \neq 0$, $F_3$=0), the threshold for $F_8$ becoming non vanishing
 appears at a large $\mu\sim $0.5 GeV as compared to the case of $F_8$=0 and $F_3 \neq $ 0.  
    
 \subsubsection{Effect of $F_3$ and $F_8$ on constituent quark masses}
\begin{figure}[!h] 
    \begin{minipage}[b]{0.5\linewidth}
    \centering
    \includegraphics[width=0.9\linewidth]{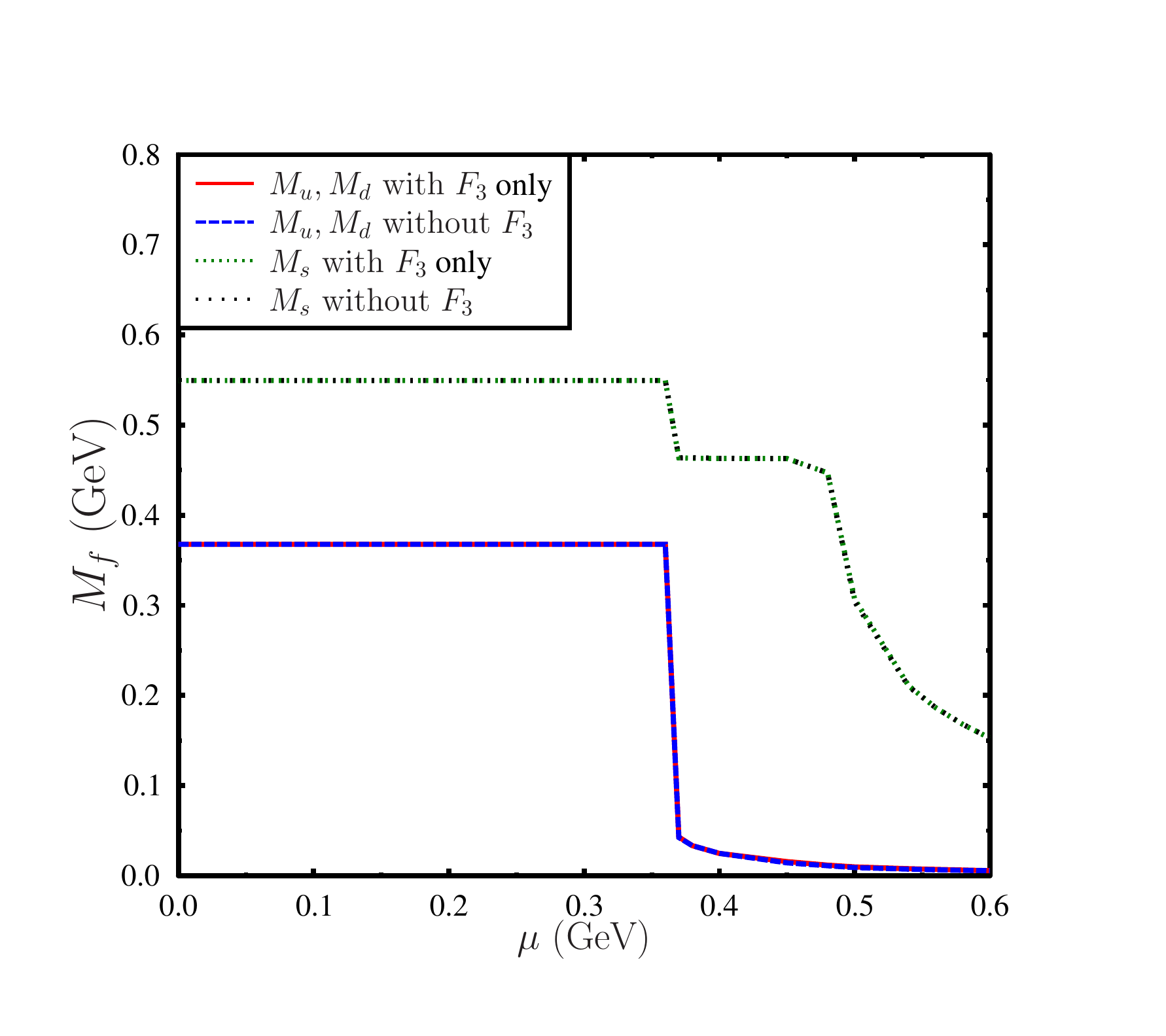} 
    \vspace{4ex}
  \end{minipage}
  \begin{minipage}[b]{0.5\linewidth}
    \centering
    \includegraphics[width=0.9\linewidth]{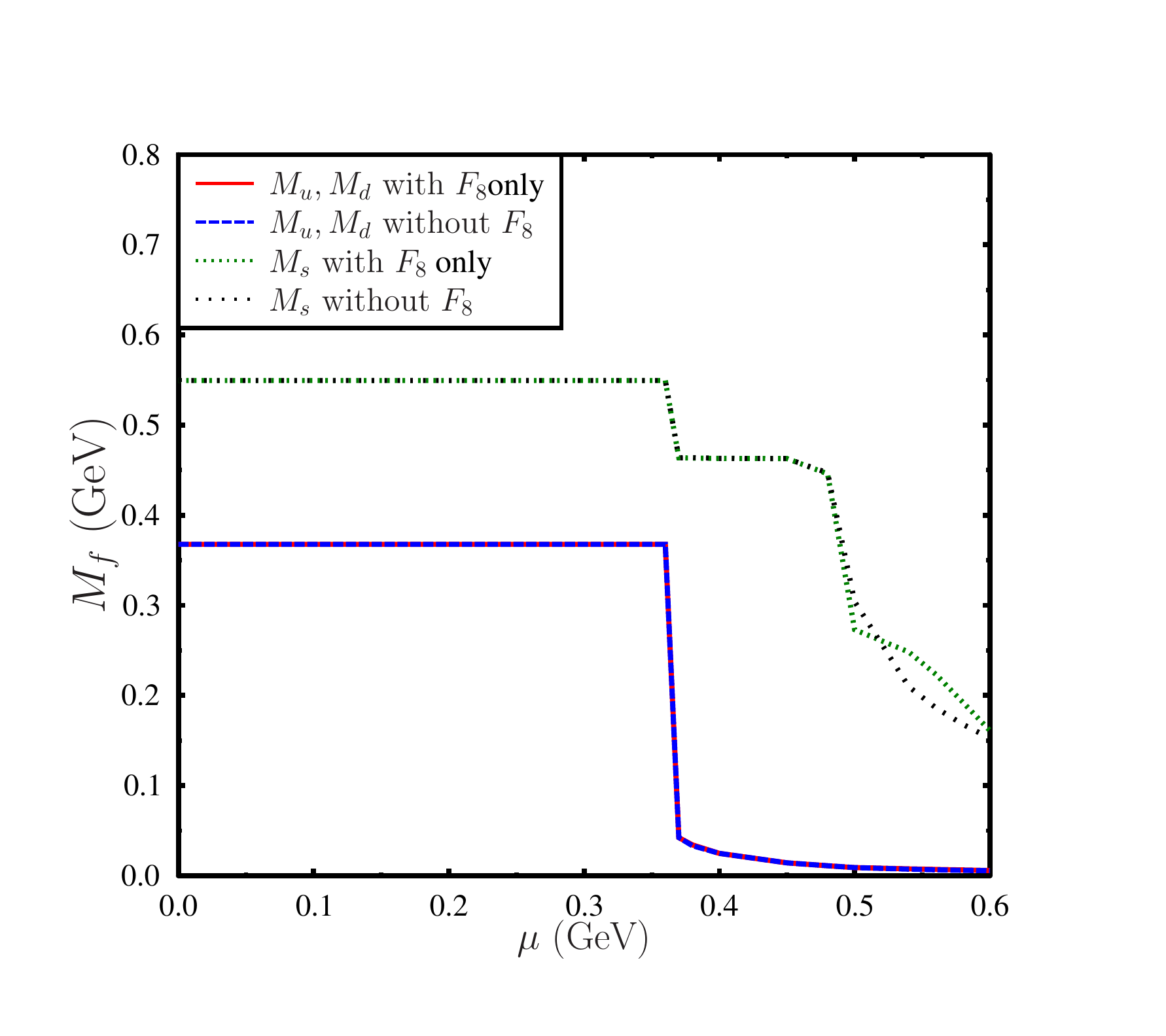} 
    \vspace{4ex}
  \end{minipage} 
\caption{Dependence of constituent quark masses as a function of the quark chemical potential at zero temperature has been shown. In the left and right 
plots we considered the effect of $F_3$ and $F_8$ respectively. In this case we have considered the tensor coupling to be $G_T=2g$. From the 
left plot it is clear that for $G_T=2g$, $F_3$ has almost no effects on the chiral phase transition as well as on the quark masses. Since $F_3$ is 
only associated with the non strange quarks, masses of the non strange quarks can be affected due to non zero spin polarization of type $F_3$. But 
in the chiral limit non strange quark masses are already very small, so the effect of $F_3$ on the masses of the non strange quark masses are
negligible. On the other hand in the right plot we can see that $F_8$ affects the strange quark mass. This is due to the fact that $F_8$ is associated
with non strange as well as strange quarks. Although in the chiral limit the masses of the non strange quarks are small but the strange quark mass
is large, hence the effect of nonzero $F_8$ on the strange quark mass can be appreciable.}
  \label{fig8}
\end{figure}

 As we have seen the behavior of $F_3$ and $F_8$ is 
 different at zero temperature as a function of chemical potential, it is also interesting to see the effect of $F_3$ and $F_8$
 on the quark masses as a function of chemical potential at zero temperature. Since $F_3$ is not associated with the strange
 quark and only $F_8$ is related to the strange quark, we can naively expect that only $F_8$ should affect the strange quark 
 mass. Indeed we can see from the Fig.\eqref{fig8} that when we consider the case $F_8=0.0$ GeV$^3$ and $F_3 \neq 0.0$  then the quark masses are almost unaffected in the presence of spin polarization. $F_3$ does affect the non strange
 scalar condensates which however is very mild on the non strange quarks already in the chiral restored phase when 
 $F_3$ is non vanishing. Since strange quark scalar condensates are affected by the non strange scalar condensates only through the determinant interaction, $F_3$ condensate has negligible effect on the strange scalar condensate as may be inferred from the Fig.\eqref{fig8}. In the opposite limit, i.e. $F_3=0.0$, $F_8 \neq 0.0$, on the other hand the strange scalar condensate gets affected by $F_8$ directly as $F_8$ appears in the dispersion relation as in Eq.\eqref{dispersion}.  
 
  It is also important to mention that near $\mu=0.55$ GeV due to the presence of $F_8$ the strange quark
 mass is slightly larger with respect to the situation when the spin polarization condensate is absent. This increase in the 
 strange quark mass is a possible artifact of the fact that spin polarization condensate also breaks the chiral symmetry and this 
 breaking of chiral symmetry possibly changes the strange quark mass. Similar behavior of the quark masses 
 have also been observed in Fig.\eqref{massfig2} for larger coupling.
 
 \subsubsection{Simultaneous $F_3$ and $F_8$}
\begin{figure}[!h] 
    \begin{minipage}[b]{0.5\linewidth}
    \centering
    \includegraphics[width=0.95\linewidth]{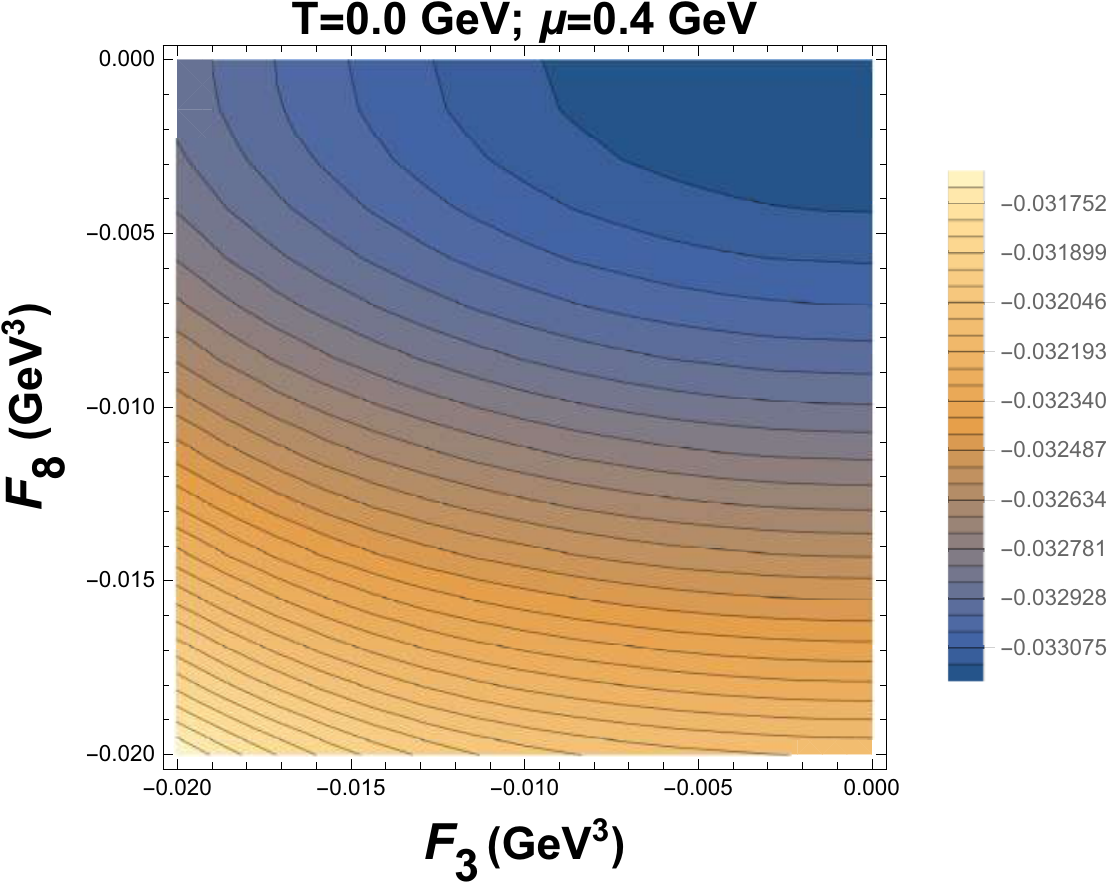} 
    \vspace{4ex}
  \end{minipage}
  \begin{minipage}[b]{0.5\linewidth}
    \centering
    \includegraphics[width=0.95\linewidth]{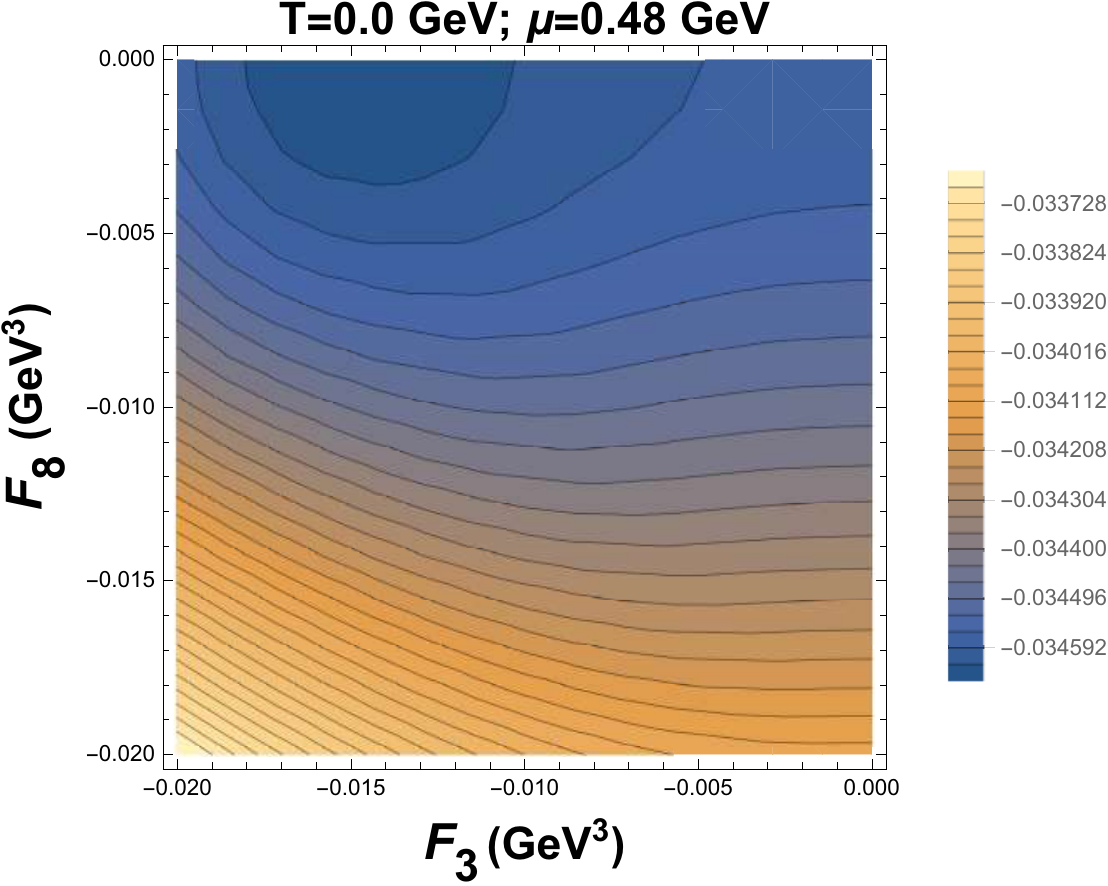} 
    \vspace{4ex}
  \end{minipage} 
  \begin{minipage}[b]{0.5\linewidth}
    \centering
    \includegraphics[width=0.95\linewidth]{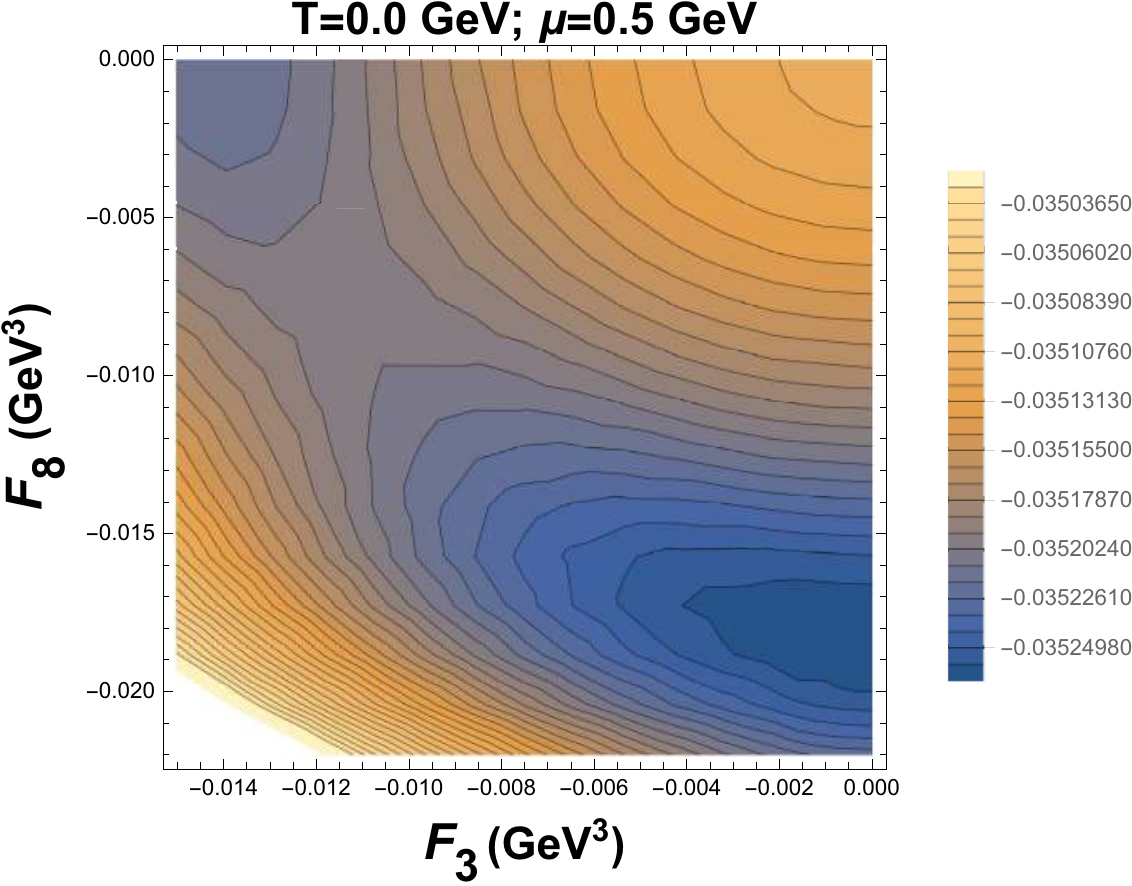} 
    \vspace{4ex}
  \end{minipage}
  \begin{minipage}[b]{0.5\linewidth}
    \centering
    \includegraphics[width=0.95\linewidth]{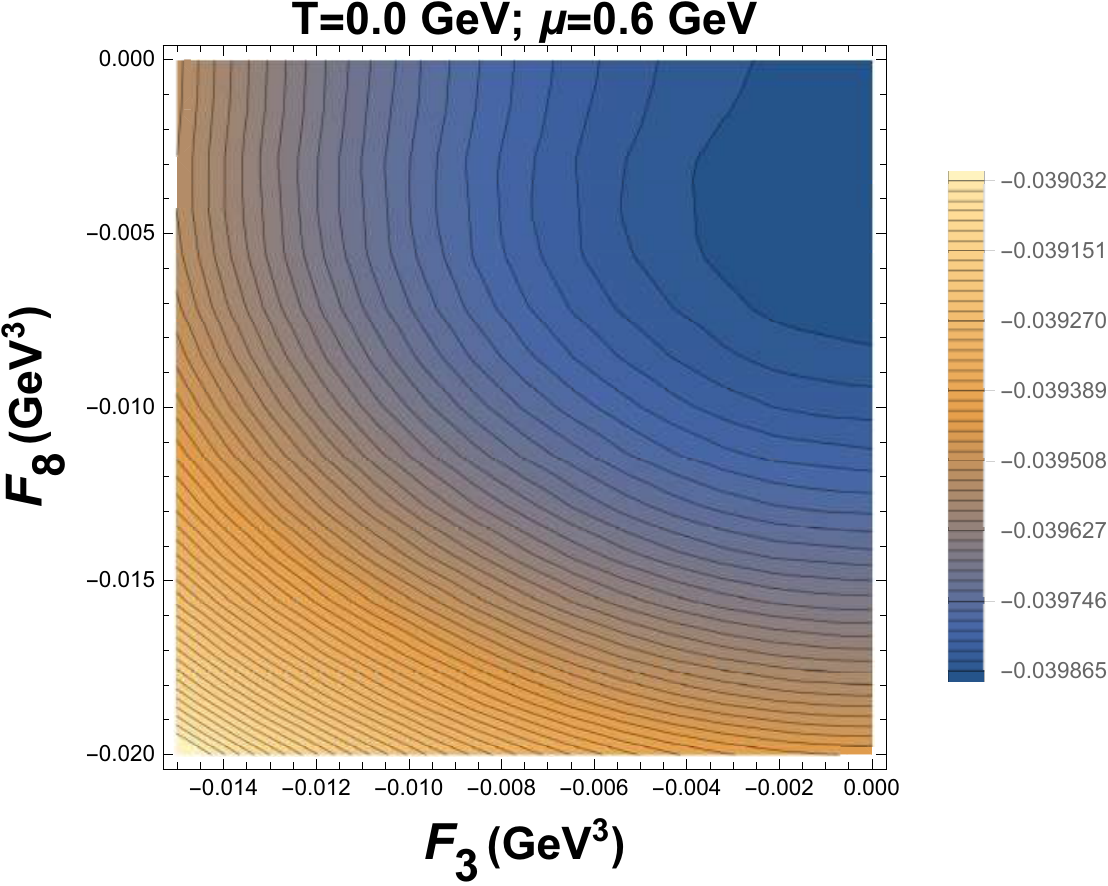} 
    \vspace{4ex}
  \end{minipage}
  \caption{This figure shows the behaviour of thermodynamic potential where $F_3$ and $F_8$ considered simultaneously in the thermodynamic
  potential at zero temperature and finite chemical potential for $G_T=2g$. This figure shows that at relatively small chemical potential 
  $\mu= 0.48$ GeV, $F_3$ develops a non zero value. However with increasing chemical potential $F_3$ melts and $F_8$ becomes non zero.
  In this figure we can see that at $\mu=0.5$GeV $F_8$ has a finite value but $F_3$ is close to zero. At a relatively higher chemical potential
  $F_3$ and $F_8$ both melt.}
  \label{newfig}
 \end{figure}

 In Fig.\eqref{newfig} we have shown the variation of $F_3$ and $F_8$ with increasing chemical potential at zero temperature. In this case 
 we have considered both $F_3$ and $F_8$ simultaneously for $G_T=2g$. It is clear from the Fig.\eqref{newfig} that non zero $F_3$ appears
 at relatively smaller $\mu$ than $F_8$. Since $F_8$ is associated with strange quark-antiquark condensate it survives even at larger 
 chemical potential relative to the $F_3$ condensate.

\subsubsection{Effect of  large tensor coupling on the chiral phase transition and quark masses}
\begin{figure}[!h] 
    \begin{minipage}[b]{0.5\linewidth}
    \centering
    \includegraphics[width=0.9\linewidth]{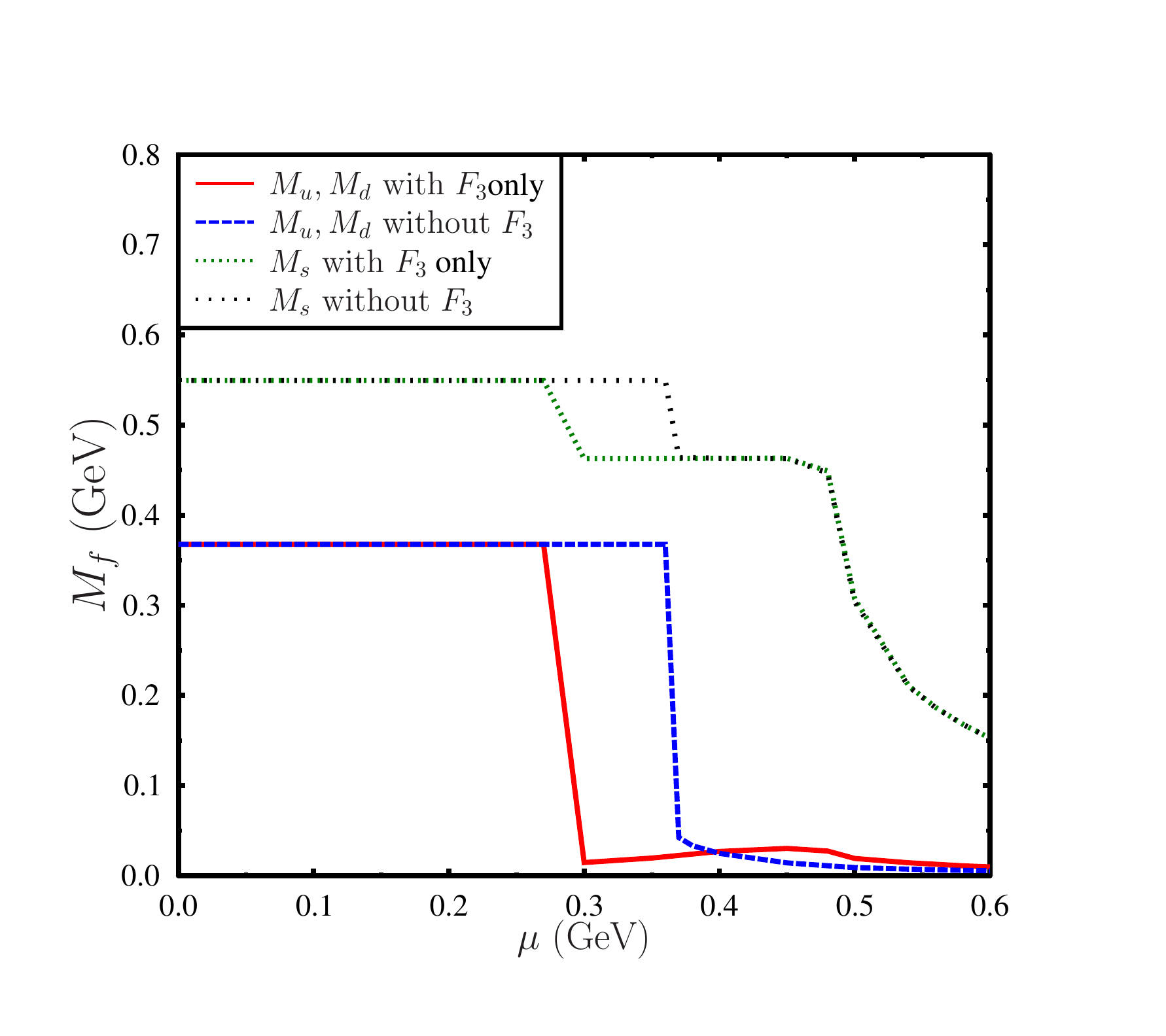} 
    \vspace{4ex}
  \end{minipage}
  \begin{minipage}[b]{0.5\linewidth}
    \centering
    \includegraphics[width=0.9\linewidth]{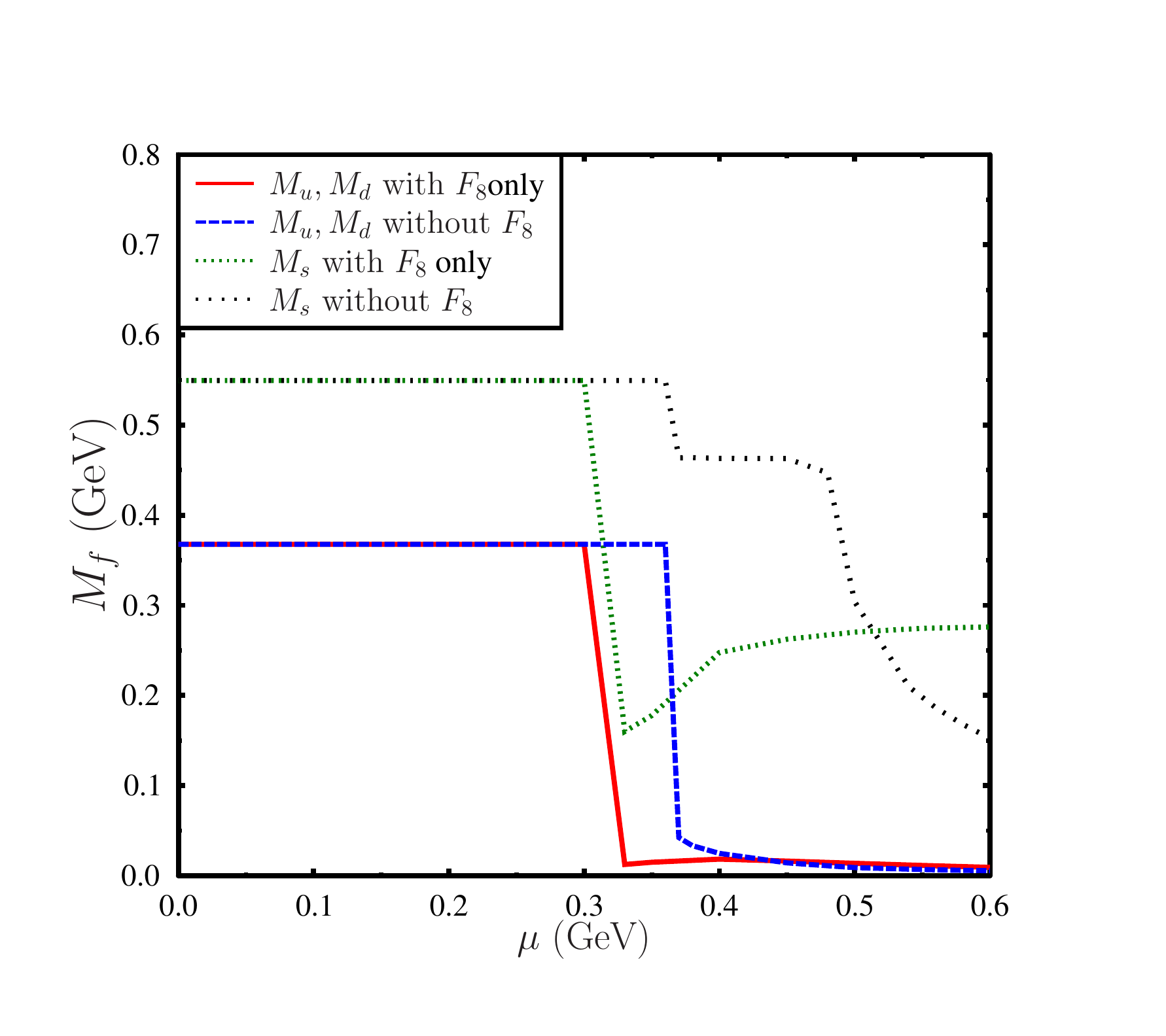} 
    \vspace{4ex}
  \end{minipage} 
\caption{This figure is similar to the Fig.\eqref{fig8}, where behavior of constituent quark masses at zero temperature and finite chemical 
potential has been shown for larger values of $G_T$. In the left plot we have considered only the effect of $F_3$ for $G_T=3.5g$ and on the 
right plot we have taken $G_T=4g$ to see the effects of $F_8$ on the chiral phase transition and the quark masses. As discussed earlier, here
also we can see that non zero $F_3$  and $F_8$ changes the critical chemical potential of chiral phase transition. However the nature of the 
chiral phase transition for $F_3$ and $F_8$ are different. The most interesting difference of $F_3$ and $F_8$ is in the behaviour of the strange quark condensate or in the strange quark mass. 
Since $F_3$ is only associated with the non strange quarks, it affects the mass of the non strange quarks directly.
Due to the flavour mixing, the strange quark mass at
the critical chemical potential changes by a small amount. But $F_8$ includes non strange as well as strange quarks, hence at the critical potential
the strange quark mass is also suffers a sudden jump. So both $\sigma_{ud}$ and $\sigma_s$ changes rapidly across the critical chemical potential.
The change in the mass of the quarks in the chiral restored phase is associated with the chiral symmetry breaking nature of the tensor spin 
polarization condensates.}
  \label{fig9}
\end{figure}

We have also checked the effects of $F_3$ and $F_8$ independently on the quark masses as well as on the chiral symmetry restoration at zero 
temperature as a function of chemical potential  for relatively larger tensor couplings. In Fig.\eqref{fig9} quark masses are plotted as 
a function of chemical potential at zero temperature considering $F_3$ and $F_8$ independently. It is important to look into 
some interesting features in the Fig.\eqref{fig9}. In the left and right plots of Fig.\eqref{fig9} we have considered the effect of $F_3$ for 
$G_T=3.5g$ and effect of $F_8$ for $G_T=4g$ respectively. As was already shown in Fig.\eqref{fig4} larger tensor coupling necessarily
changes the critical chemical potential for chiral transition, Fig.\eqref{fig9} also shows this behaviour. However it is interesting to point out the 
the crucial difference in the chiral symmetry breaking pattern in this case. In the left plot of Fig.\eqref{fig9} where only the effect of $F_3$
has been considered we can see that at the critical chemical potential non-strange quark masses changes steeply, indicating a first order chiral 
phase transition. However the strange quark mass changes slightly due to the fact that determinant interacting term generate flavour mixing terms. 
Since $F_3$ is only associated with the non strange quark flavour this behavior of the quark mass is expected, where the nonstrange quark 
masses are affected primarily and the strange quark mass is affected due to the flavour mixing determinant interaction. On the other hand, the right plot in Fig.\eqref{fig9}
shows the behavior of the quark masses as a function of chemical potential for nonzero $F_8$ only, i.e. $F_3=0.0, F_8 \neq 0.0$. In this plot, one can clearly see the 
effect of $F_8$ on the quark masses particularly on the strange quark mass. Since $F_8$ is also associated with the strange quarks, apart from the 
non strange quarks it directly affects the strange quark condensate and hence the strange quark mass. This behavior of the strange quark mass at 
the critical chemical potential indicates that the chiral symmetry restoration for a large tensor coupling, where the spin polarization plays 
a important role  and infact acts as a catalyst of chiral symmetry restoration. 

\section{Conclusions}
\label{conclusion}
In this work, we have considered the 2+1 flavor NJL model in the presence of tensor interaction with non zero current quark masses.
The original idea of the presence of spin polarization in quark liquid was motivated considering one gluon exchange interactions
in perturbative QCD processes \cite{Tatsumi2000}. Ferromagnetic quark matter can arise due to both axial vector 
and tensor type interaction.
Although the axial vector type interaction can be generated from the one gluon exchange QCD interaction by Fierz transformation, the tensor
type interactions cannot be generated using Fierz transformation. Thus at very high densities where perturbative QCD processes are relevant,
tensor type of interaction will not be suitable to study spin polarization in quark matter. More importantly at moderate densities close to the chiral
phase transition one expects nonperturbative effects to play an important role. In the present investigation within the ambit of NJL model applied to moderate densities, we have considered only the tensor type four point interaction. 
We might note here that the coupling constant of the tensor interaction is related to the scalar and pseudo scalar channel. However in general,  this tensor coupling constant can be independent. We take the coupling constant of the tensor interaction $G_T$
as a parameter of the model. We have taken various values of the tensor couplings $G_T$, e.g. $G_T=2.0g$ and lower as well as relatively
larger values of $G_T$, e.g. $G_T=4g, 3.5g$ etc.\\

 For 2+1 flavor NJL model, tensor type interaction at the mean field level leads to two types of spin polarization 
 condensates, $F_3=\langle \bar{\psi} \Sigma_z \lambda_3 \psi \rangle$ and $F_8=\langle \bar{\psi} \Sigma_z \lambda_8
 \psi \rangle$. Since we have various condensates in 2+1 flavor NJL model in the presence of tensor interaction
 we take a rather simplified approximation, where $F_3$ and $F_8$ are not independent rather $F_8=F_3/\sqrt{3}$. 
 One may note that in general $F_3$ and $F_8$ are independent due to the fact that $F_8$ is associated with the 
 strange quark spin polarization condensate, on the other hand $F_3$ contains only $u, d$ quark spin polarization 
 condensates. Therefore we have also considered the case where $F_3$ and $F_8$ are treated independently.  Generically spin polarization  for moderate 
 tensor coupling (e.g. $G_T=2g$) does not appear at zero temperature and zero chemical potential,
 rather it appears at high $\mu$ in the chiral restored phase.  
 At large chemical potential and small temperature the generic feature of such spin polarized condensate
 lies in affecting the strange quark mass rather than the non-strange quark masses for moderate tensor coupling. 
 Such spin polarized condensate vanishes for temperatures of the order of few tens of MeV and thus can be relevant for 
 neutron stars and proto neutron stars. We also find that there is a threshold tensor coupling, below which the spin 
 polarization condensates do not develop.\\ 
  
  For larger tensor coupling (e.g. $G_T=4g$) it is observed that the magnitude of the spin polarized condensate is larger
  and it also affects the critical  chemical potential for chiral symmetry restoration. In fact such condensate catalyzes the chiral restoration
  in the sense that $\mu_c$ is small in presence of spin polarization as compared to the case when such a condensate
  is not there.\\ 
  
  Unlike superconducting diquark condensate, the spin polarization condensate is not a monotonic function of chemical
  potential and as the chemical potential is increased the magnitude becomes a maximum beyond which it vanishes when
  $\mu$ is increased further. The range of chemical potential for which such condensate exists as well as the magnitude of 
  the condensate, increases with the strength of the tensor coupling.
  

 \section*{Acknowledgement}

 RKM would like to thank PRL for support and local hospitality for her visit, during which this problem was initiated. Also
 RKM would like to thank Basanta K. Nandi and Sadhana Dash for constant support and encouragement.

\end{document}